\documentstyle[preprint,eqsecnum,aps]{revtex}
\tighten

\newcommand{\beq}{\begin{equation}}
\newcommand{\eeq}{\end{equation}}
\newtheorem{theorem}{Theorem}

\begin{document}
\draft
\title{Topology of the gauge-invariant gauge field in two-color QCD.}
\author{Kurt Haller\thanks{Department of Physics;~e-mail: khaller@uconnvm.uconn.edu}, 
Lusheng Chen\thanks{Department of Physics;~e-mail: chen@phys.uconn.edu}and 
Y. S. Choi\thanks{Department of Mathematics;~e-mail: choi@math.uconn.edu}}
\address{University of Connecticut, Storrs, Connecticut 06269}
\maketitle
\begin{abstract}
We investigate solutions to a nonlinear integral equation which has a central role in implementing the 
non-Abelian Gauss's Law and in constructing gauge-invariant quark and gluon fields. Here we concern 
ourselves with solutions to this same equation that are not operator-valued, but are functions of 
spatial variables and carry spatial and SU(2) indices. 
We obtain an expression for the gauge-invariant gauge field in two-color QCD, 
define an index that we will refer to as
the ``winding number'' that
characterizes it, and show that this winding number is invariant to a small gauge transformation
of the gauge field on which our construction of the gauge-invariant gauge field is based. 
We discuss the role of this gauge field in determining the winding number of 
the gauge-invariant gauge field. We also show that when the winding number
of the gauge field is an integer $\ell{\neq}0$, the gauge-invariant gauge field manifests 
winding numbers
that are not integers, and are half-integers only when $\ell=0$.
\end{abstract}

\section{Introduction}
\label{sec-Int}
In earlier work, in which we discussed QCD in the temporal ($A_i^{\gamma}=0)$ gauge,\cite{CBH2} 
we implemented the non-Abelian Gauss's law and constructed 
 quark and gluon operator-valued fields that are invariant to small non-Abelian gauge transformations. 
An essential element in that 
program was a nonlocal operator, $\overline{{\cal{A}}_{i}^{\gamma}}({\bf{r}})$, which we 
called the ``resolvent field'' in later publications.\cite{CHICHEP,Hparis} The resolvent field is a functional
of the gauge field $A_i^{\gamma}({\bf{r}})$, and it has a pivotal role in our work.
It first appears, in combination with ${\Pi}_i^{\gamma}({\bf{r}})$ --- the canonical momentum for the gauge 
field and the negative chromoelectric field --- in the operator, 
\begin{equation}
{\Psi}=||{\exp}\left(i{\int}d{\bf r}\overline{{\cal{A}}}_{i}^{\gamma}({\bf{r}})
{\Pi}_i^{\gamma}({\bf{r}})\right)||,
\label{eq:Psi}
\end{equation}
where the double-bar bracket denotes a normal ordering in which all the gauge fields are 
placed to the left of all canonical momenta, so that when the exponential is expanded, all powers of 
$\overline{{\cal {A}}}_{i}^{\gamma}({\bf r})$ appear to the left of any of the  ${\Pi}_i^{\gamma}({\bf{r}})$.
${\Psi}$ converts any state $|\phi\rangle$ that obeys the Abelian Gauss's law 
${\partial}_i{\Pi}_i^{\alpha}({\bf{r}})|\phi\rangle=0$ into a state ${\Psi}|\phi\rangle$ that obeys the 
``pure-glue'' non-Abelian Gauss's law $[{\partial}_i{\Pi}_i^{\alpha}({\bf{r}}) +gf^{\alpha\beta\gamma}
A_i^{\beta}({\bf{r}}){\Pi}_i^{\gamma}({\bf{r}})]{\Psi}|\phi\rangle=0\,.$
\footnote{We use non-relativistic notation in this work, in which subscripted indices 
designate contravariant components of contravariant quantities, such as $A^{\gamma}_i$, 
and covariant components of covariant quantities, such as $\partial_i$. 
This convention is used extensively throughout this work.} \bigskip

The resolvent field reappears, in the form  
\begin{equation}
\overline{{\cal Y}^{\alpha}}({\bf r})=
{\textstyle \frac{\partial_{j}}{\partial^{2}}
\overline{{\cal A}_{j}^{\alpha}}({\bf r})}\,,
\label{eq:Y}
\end{equation}
in the expression for the gauge-invariant quark field
\begin{equation}
\psi_{\sf GI}({\bf{r}})=V_{\cal{C}}({\bf{r}})\,\psi({\bf{r}})
\label{eq:GIpsi}
\end{equation}
where 
\begin{equation}
V_{\cal{C}}({\bf{r}})=\exp\left(\,-ig{\overline{{\cal{Y}}^\alpha}}({\bf{r}})
{\textstyle\frac{\lambda^\alpha}{2}}\,\right)\,
\exp\left(-ig{\cal X}^\alpha({\bf{r}})
{\textstyle\frac{\lambda^\alpha}{2}}\right)
\label{eq:VC}
\end{equation}
and 
\begin{equation}
{\cal{X}}^\alpha({\bf{r}}) =
[\,{\textstyle\frac{\partial_i}{\partial^2}}A_i^\alpha({\bf{r}})\,]\,.
\end{equation}
The resolvent field also appears in the expression for the gauge-invariant gluon field 
\begin{equation}
[\,A_{{\sf GI}\,i}^{{\gamma}}({\bf{r}})\,{\textstyle\frac{{\lambda}^{\gamma}}{2}}\,]
=V_{\cal{C}}({\bf{r}})\,[\,A_{i}^{\gamma}({\bf{r}})\,
{\textstyle\frac{\lambda^\gamma}{2}}\,]\,
V_{\cal{C}}^{-1}({\bf{r}})
+{\textstyle\frac{i}{g}}\,V_{\cal{C}}({\bf{r}})\,
\partial_{i}V_{\cal{C}}^{-1}({\bf{r}})\,,
\label{eq:AdressedAxz}
\end{equation}
which can also be expressed as
\begin{equation}
A_{{\sf GI}\,i}^{\gamma}({\bf{r}})=
A_{T\,i}^{\gamma}({\bf{r}}) +
\left[\delta_{ij}-{\textstyle\frac{\partial_{i}\partial_j}
{\partial^2}}\right]\,\overline{{\cal{A}}^{\gamma}_{j}}({\bf{r}})\,.
\label{eq:GIglue}
\end{equation}

In subsequent work,\cite{CHICHEP,Hparis,BCH3,CHQC} we used these gauge-invariant fields to 
derive an expression
 for a nonlocal interaction that couples gauge-invariant quark color-charge densities to each other 
and to gluons through ``chains'', in which each link consists of an 
integral over the gauge-invariant gluon field. SU(3) or SU(2) structure constants 
connect the 
links to form the chains. As was shown in Ref. \cite{CBH2}, 
the quantities that appear in this formulation --- the gauge-invariant quark color-charge density 
$j^{\alpha}_{0\;{\sf GI}}=g{\psi}^{\dagger}_{\sf GI}\frac{\lambda^{\alpha}}{2}{\psi}_{\sf GI}$ and
the  gauge-invariant  gauge field $A_{{\sf GI}\,i}^{\gamma}({\bf{r}})$ --- commute with the 
generator of infinitesimal gauge transformations and therefore are completely invariant to small gauge 
transformations. In the work presented here, we will focus attention on the  resolvent field in two-color QCD. 
We will examine  its structure in configuration space and its role in determining the 
topological features of the gauge-invariant gluon field. 

\section{Configurations of the Resolvent Field}
\label{sec-config}
In Ref.\cite{CBH2}, we presented a nonlinear integral equation for the resolvent field. 
Although, in Ref.\cite{CBH2}, our concern was with the series representation of the operator-valued 
resolvent field, it is also relevant to ask whether the integral equation for the resolvent field
has solutions other than that series. It is this question that we now consider.
The most general form of the 
integral equation for the resolvent field, applicable to any SU(N) gauge theory, 
is not useful for this 
investigation; but its SU(2) 
form, given in Ref. \cite{CBH2}, is very suitable for our purposes. 
In the SU(2) case, the integral equation for the 
resolvent field
$\overline{{\cal{A}}_{i}^{\gamma}}({\bf{r}})$ can be expressed in the following form:
\begin{equation}
\overline{{\cal{A}}_{i}^{\gamma}}({\bf{r}})=
\overline{{\cal{A}}_{i}^{\gamma}}({\bf{r}})_{\cal X}+
\overline{{\cal{A}}_{i}^{\gamma}}({\bf{r}})_{\overline{\cal Y}}\;,
\label{eq:azero}
\end{equation}
where
\begin{eqnarray}
\overline{{\cal{A}}_{i}^{\gamma}}({\bf{r}})_{\cal X}
&&\,=g\epsilon^{\alpha\beta\gamma}\,
{\cal{X}}^\alpha({\bf{r}})\,A_i^\beta({\bf{r}})\,
{\textstyle\frac{\sin({\cal{N}})}{{\cal{N}}}}
\nonumber\\
&&-g\epsilon^{\alpha\beta\gamma}\,
{\cal{X}}^\alpha({\bf{r}})\,\partial_i{\cal{X}}^\beta({\bf{r}})\,
{\textstyle\frac{1-\cos({\cal{N}})}{{\cal{N}}^2}}
\nonumber\\
&&-g^2\epsilon^{\alpha\beta b}
\epsilon^{b\mu\gamma}\,
{\cal{X}}^\mu({\bf{r}})\,{\cal{X}}^\alpha({\bf{r}})
\,A_i^\beta({\bf{r}})\,
{\textstyle\frac{1-\cos({\cal{N}})}{{\cal{N}}^2}}
\nonumber\\
&&+g^2\epsilon^{\alpha\beta b}
\epsilon^{b\mu\gamma}\,
{\cal{X}}^\mu({\bf{r}})\,{\cal{X}}^\alpha({\bf{r}})\,
\partial_i{\cal{X}}^\beta({\bf{r}})\,
[{\textstyle\frac{1}{{\cal{N}}^2}}
-{\textstyle\frac{\sin({\cal{N}})}{{\cal{N}}^3}}]
\label{eq:a2X}
\end{eqnarray}
and
\begin{eqnarray}
\overline{{\cal{A}}_{i}^{\gamma}}({\bf{r}})_{\overline{\cal Y}}
&&\;=g\epsilon^{\alpha\beta\gamma}\,
\overline{{\cal{Y}}^\alpha}({\bf{r}})
\,\left(\,A_{T\,i}^\beta({\bf{r}}) +
(\delta_{ij}-{\textstyle\frac{\partial_{i}\partial_j}
{\partial^2}})\overline{{\cal{A}}^\beta_{j}}({\bf{r}})\,\right)\,
{\textstyle\frac{\sin(\overline{\cal{N}})}{\overline{\cal{N}}}}
\nonumber\\
&&+g\epsilon^{\alpha\beta\gamma}\,
\overline{{\cal{Y}}^\alpha}({\bf{r}})
\,\partial_i\overline{{\cal{Y}}^\beta}({\bf{r}})\,
{\textstyle\frac{1-\cos(\overline{\cal{N}})}
{\overline{\cal{N}}^2}}
\nonumber\\
&&+g^2\epsilon^{\alpha\beta b}
\epsilon^{b\mu\gamma}\,
\overline{{\cal{Y}}^\mu}({\bf{r}})\,
\overline{{\cal{Y}}^\alpha}({\bf{r}})
\,\left(\,A_{T\,i}^\beta({\bf{r}}) +
(\delta_{ij}-{\textstyle\frac{\partial_{i}\partial_j}
{\partial^2}})\overline{{\cal{A}}^\beta_{j}}({\bf{r}})\,\right)\,
{\textstyle\frac{1-\cos(\overline{\cal{N}})}{\overline{\cal{N}}^2}}
\nonumber\\
&&+g^2\epsilon^{\alpha\beta b}
\epsilon^{b\mu\gamma}\,
\overline{{\cal{Y}}^\mu}({\bf{r}})\,
\overline{{\cal{Y}}^\alpha}({\bf{r}})\,
\partial_i\overline{{\cal{Y}}^\beta}({\bf{r}})\,
[{\textstyle\frac{1}{\overline{\cal{N}}^2}}-
{\textstyle\frac{\sin(\overline{\cal{N}})}
{\overline{\cal{N}}^3}}]\;.
\label{eq:a2Y}
\end{eqnarray}
 We observe that 
the right-hand side of Eq.(\ref{eq:a2X}) is independent of 
$\overline{{\cal{A}}_{i}^{\gamma}}({\bf{r}})$ and depends only on 
the gauge field $A_i^{\gamma}({\bf{r}})$ and on trigonometric functions of 
${\cal{N}}$, where
\begin{equation}
{\cal{N}}=\left[g^2\,
{\cal{X}}^\alpha({\bf{r}})\,
{\cal{X}}^\alpha({\bf{r}})\,\right]^{\frac{1}{2}}\;.
\label{eq:defN}
\end{equation} 
$\overline{{\cal{A}}_{i}^{\gamma}}({\bf{r}})_{\cal X}$ therefore is not responsible for any of the 
nonlinearity of the nonlinear integral equation shown in Eq.(\ref{eq:azero}), but represents an 
inhomogeneity in that equation. 
The right-hand side of Eq.(\ref{eq:a2Y}), however, contains the resolvent field 
$\overline{{\cal{A}}_{i}^{\gamma}}({\bf{r}})$ and $\overline{\cal{N}}$ as well as 
trigonometric functions of 
$\overline{\cal{N}}$, where 
\begin{equation}
\overline{\cal{N}}=
\left[g^2\,\overline{{\cal{Y}}^\alpha}({\bf{r}})\,
\overline{{\cal{Y}}^\alpha}({\bf{r}})\,\right]^{\frac{1}{2}}\,.
\label{eq:defNbar}
\end{equation} 
$\overline{{\cal{A}}_{i}^{\gamma}}({\bf{r}})_{\overline{\cal Y}}$ therefore is a nonlinear
functional of the resolvent field
$\overline{{\cal{A}}_{i}^{\gamma}}({\bf{r}})$, and of $\overline{{\cal{Y}}^\gamma}({\bf{r}})$,
whose dependence on $\overline{{\cal{A}}_{i}^{\gamma}}({\bf{r}})$ is given in Eq.(\ref{eq:Y}). 
It is the structure of $\overline{{\cal{A}}_{i}^{\gamma}}({\bf{r}})_{\overline{\cal Y}}\,$ that
makes Eq.(\ref{eq:azero}) a nonlinear integral equation. \bigskip

Eqs.(\ref{eq:GIpsi}) and (\ref{eq:AdressedAxz}) are identical in form
 to the equations describing the change in a spinor field and a gauge field, respectively,  
produced by a finite non-Abelian gauge transformation, with the important exception that 
the transforming $c$-number gauge function in a gauge transformation is replaced by the 
operator-valued  $g\overline{\cal{Y}}^\alpha({\bf{r}})$ and $g{\cal{X}}^\alpha({\bf{r}})$ in
$V_{\cal{C}}({\bf{r}})$. The SU(2) version of $V_{\cal{C}}({\bf{r}})$, given by
\begin{equation}
\left[V_{\cal{C}}({\bf{r}})\right]_{SU(2)}=\exp\left(\,-ig{\overline{{\cal{Y}}^\alpha}}({\bf{r}})
{\textstyle\frac{\tau^\alpha}{2}}\,\right)\,
\exp\left(-ig{\cal X}^\beta({\bf{r}})
{\textstyle\frac{\tau^\beta}{2}}\right)\,,
\label{eq:VCSU2}
\end{equation} 
can be interpreted as a sequence of two rotations represented in the 
fundamental representation of SU(2). In 
one of these rotations, $g{\cal{X}}^\alpha({\bf{r}})/{\cal N}$ designates 
a direction about which the  rotation is made, and 
${\cal N}$ the magnitude of the locally varying angle of rotation; a similar correspondence
applies to $g\overline{\cal{Y}}^\alpha({\bf{r}})$ and $\overline{\cal{N}}$ in the other rotation.
In this SU(2) case, we can also express 
$\left[V_{\cal{C}}({\bf{r}})\right]_{SU(2)}$ in the form 
\begin{equation}
\left[V_{\cal{C}}({\bf{r}})\right]_{SU(2)}=\exp\left(\,-ig{\overline{{\cal{Z}}^\alpha}}({\bf{r}})
{\textstyle\frac{\tau^\alpha}{2}}\,\right)
\label{eq:VCSU2Z}
\end{equation}
where the composition laws for rotations enable us
to express ${\overline{{\cal{Z}}^\alpha}}({\bf{r}})$ as a function of ${{{\cal{X}}^\alpha}}({\bf{r}})$
and ${\overline{{\cal{Y}}^\alpha}}({\bf{r}})$, and both in terms of equivalent sets of locally varying 
Euler angles, much as was done by Christ and Lee.\cite{christlee}
Because the algebra of SU(N) generators is closed, the SU(3) form of 
$V_{\cal{C}}({\bf{r}})$ can also be expressed
in the form $V_{\cal{C}}({\bf{r}})=
\exp\left(\,-ig{\overline{{\cal{Z}}^\alpha}}({\bf{r}})
{\textstyle\frac{\lambda^\alpha}{2}}\,\right),$ where the Baker-Hausdorff-Campbell 
theorem enables us to equate 
$\overline{{\cal{Z}}^\alpha}$ to a complicated expansion in ${{\cal{X}}^\alpha}$
and $\overline{{\cal{Y}}^\alpha}$. 
The familiar group
composition laws for rotations are a special 
case of the Baker-Hausdorff-Campbell theorem applied to the SU(2) case.
Eqs.(\ref{eq:a2X}) and $(\ref{eq:a2Y})$ illustrate this similarity in form 
of Eqs.(\ref{eq:GIpsi}) and (\ref{eq:AdressedAxz}) to 
the equations that implement finite gauge transformations.
Eq.(\ref{eq:a2X}) has the same form as a finite gauge transformation of a gauge field in the 
adjoint representation of SU(2), with $g{\cal{X}}^\alpha({\bf{r}})$ serving as the 
function by which the field is gauge-transformed; 
the same applies to $(\ref{eq:a2Y})$ with $g\overline{\cal{Y}}^\alpha({\bf{r}})$ 
in place of $g{\cal{X}}^\alpha({\bf{r}})$.
Such non-Abelian gauge configurations have well-documented important geometric and 
topological implications for gauge theories.\cite{Jack,Rho,Bhad,Zahed}
The similarity in form of the equations that establish gauge-invariant fields to those that 
effect gauge transformations accounts for the appearance in our work of 
expressions that resemble the inverse of the Faddeev Popov operator 
$\partial_i({\delta}_{ac}\partial_i+gf^{cba}A_i^b)$, which arises when transformations from the 
temporal to the Coulomb gauge are carried out by treating them as transformations from Cartesian to 
curvilinear field variables.\cite{christlee,creutz,sakita} In 
our work, $\partial_i({\delta}_{ac}\partial_i+gf^{cba}A_{{\sf GI}\,i}^b\,)$
appears in place of the Faddeev-Popov operator, 
and it appears in a representation in which the gauge field $A_{{\sf GI}\,i}^b$ and 
the quark color charge density $j_0^b=g{\psi}^{\dagger}({\lambda}^b/2){\psi}$ are 
gauge invariant and in which the temporal gauge condition continues to apply, even 
though the gauge-invariant gauge field (but not the gauge field) is transverse.~\cite{BCH3,CHQC}
\bigskip

In spite of this similarity in form of finite gauge transformations to
Eqs.(\ref{eq:GIpsi}) and (\ref{eq:AdressedAxz}), the significance of 
these two equations is very different from that of a finite gauge 
transformation. In a gauge transformation, the gauge field is 
transformed by an arbitrary $c$-number function that bears no relationship 
to the original gauge field. But in the case of Eqs.(\ref{eq:GIpsi}) and (\ref{eq:AdressedAxz}), 
 ${{{\cal{X}}^\alpha}}$ and ${\overline{{\cal{Y}}^\alpha}}$ (as well as ${\cal{N}}$ and
$\overline{\cal{N}}$) are themselves functionals of the gauge field, and therefore are subject to 
the same gauge transformations that affect it. In fact, under the infinitesimal gauge transformation\
${\delta}A_i^{\gamma}({\bf r})={\partial}_i{\delta}{\omega}^{\gamma}({\bf r})+
g{\epsilon}_{{\gamma}{\alpha}{\beta}}
A_i^{\alpha}({\bf r}){\delta}{\omega}^{\beta}({\bf r})\,,$ $V_{\cal{C}}({\bf{r}})$ transforms so that
${\delta}V_{\cal{C}}({\bf{r}})=-igV_{\cal{C}}({\bf{r}}){\delta}{\omega}^{\gamma}({\bf r})
\frac{{\lambda}^{\gamma}}{2}$.\cite{mario}  Instead of representing a 
gauge transformation, Eqs.(\ref{eq:GIpsi}) and (\ref{eq:AdressedAxz}) establish the operator-valued 
nonlocal quantities $\psi_{\sf GI}({\bf{r}})$ and $A_{{\sf GI}\,i}^{{\gamma}}({\bf{r}})$, 
which are {\em entirely invariant} to any further transformations effected by the 
generator of gauge transformations 
\begin{equation}
{\sf T}_G={\exp}\left[i{\int}d{\bf{r}}\left(\partial_i{\Pi}^{\gamma}_i({\bf{r}})+gf^{\gamma\beta\alpha}
A_i^{\beta}({\bf{r}}){\Pi}_i^{\alpha}({\bf{r}})+g{\psi}^{\dagger}({\bf{r}})
\frac{{\lambda}^{\gamma}}{2}{\psi}({\bf{r}})\right)
{\omega}^{\gamma}({\bf{r}})\right].
\label{eq:GTop}
\end{equation}

The integral equation described in Eqs.(\ref{eq:azero})-(\ref{eq:a2Y}) can be viewed in two distinct ways: 
On the one hand, as was shown in Ref.\cite{CBH2}, Eqs.(\ref{eq:azero})-(\ref{eq:a2Y}) 
can be used to generate an operator-valued series 
representation of the resolvent field $\overline{{\cal{A}}_{i}^{\gamma}}({\bf{r}})$, which is 
instrumental in establishing gauge-invariant quark and gluon fields and quantum states that implement 
the non-Abelian Gauss's law. But, because all the operator-valued fields that appear in it commute with 
each other (the only operator that would not commute with the quantities in Eqs.(\ref{eq:azero})-(\ref{eq:a2Y}) 
--- $\Pi^\alpha_i({\bf r})$, 
the momentum conjugate to $A^\alpha_i({\bf r})$ --- does not appear in any of these equations), 
 Eq.(\ref{eq:azero}) also can be treated as a non-linear integral equation whose solutions 
can be investigated by more-or-less standard procedures. 
In this second context, we can view the resolvent field 
$\overline{{\cal{A}}_{i}^{\gamma}}({\bf{r}})$ and the gauge field $A_i^{\gamma}({\bf r})$ as functions of 
spatial variables that obey the same nonlinear integral equation as the 
corresponding operator-valued quantities. We can, furthermore, make an {\em ansatz} about the functional 
dependence of $\overline{{\cal{A}}_{i}^{\gamma}}({\bf{r}})$ and $A_i^{\gamma}({\bf r})$ 
on spatial variables that is 
physically reasonable, and that facilitates the solution of the nonlinear equation given in 
Eqs.(\ref{eq:azero})-(\ref{eq:a2Y}). The functional forms of  $\overline{{\cal{A}}_{i}^{\gamma}}({\bf{r}})$ and 
$A_i^{\gamma}({\bf r})$ will represent number-valued realizations of these operator-valued fields, 
and will enable us to investigate their spatial configurations and their topological features. 
\bigskip

We will carry out this program by assuming that
$\overline{{\cal{A}}_{i}^{\gamma}}({\bf{r}})$ and 
$A_i^{\gamma}({\bf r})$ are functions of spatial variables, that they 
are second-rank tensors 
in the combined spatial and SU(2) indices $i$ and $\gamma$ respectively, and that, except in so far as 
the forms of $\overline{{\cal{A}}_{i}^{\gamma}}({\bf{r}})$ and 
$A_i^{\gamma}({\bf r})$ must reflect this second-rank 
tensor structure, they are isotropic functions of the position. The most general form of 
$\overline{{\cal{A}}_{i}^{\gamma}}({\bf{r}})$ will therefore be 
\begin{equation}
\overline{{\cal{A}}_{i}^{\gamma}}({\bf{r}})={\delta}_{{i}\,{\gamma}}\,{\xi}_{A}(r)+
\frac{r_i\,r_{\gamma}}{r^2}\,{\xi}_{B}(r)+{\epsilon}_{i{\gamma}n}\frac{r_n}{r}\,{\xi}_{C}(r)
\label{eq:xistruct}
\end{equation}
where ${\xi}_{A}(r)$, ${\xi}_{B}(r)$, and ${\xi}_{C}(r)$ represent as yet unspecified isotropic functions of $r$.
It is convenient to separate the resolvent field 
$\overline{{\cal {A}}_i^{\gamma}}({\bf r})$ into longitudinal and 
transverse parts, $\overline{{\cal {A}}_i^{\gamma}}\,^L({\bf r})$ and 
$\overline{{\cal {A}}_i^{\gamma}}\,^T({\bf r})$ respectively. We then use the representations 
\begin{equation} 
\overline{{\cal {A}}_i^{\gamma}}\,^L({\bf r})={\delta}_{{i}\,{\gamma}}\,{\Phi}(r)+
\frac{r_i\,r_{\gamma}}{r}\,{\Phi}^{\prime}(r)\;\;\;\mbox{where}\;\;{\Phi}^{\prime}(r)=\frac{d\Phi}{dr}
\label{eq:longA}
\end{equation}
and 
\begin{equation}
\overline{{\cal {A}}_i^{\gamma}}\,^T({\bf r})={\delta}_{{i}\,{\gamma}}\,{\varphi}_{A}(r)+
\frac{r_i\,r_{\gamma}}{r^2}\,{\varphi}_{B}(r)+{\epsilon}_{i{\gamma}n}\frac{r_n}{r}\,{\varphi}_{C}(r)
\label{eq:transA}
\end{equation}
where the transversality of $\overline{{\cal {A}}_i^{\gamma}}\,^T({\bf r})$ requires that
\begin{equation}
\frac{d(r^2{\varphi}_B)}{dr}+r^2\frac{d\,\varphi_A}{dr}=0\,.
\label{eq:transcond}
\end{equation}
The transversality of $\overline{{\cal {A}}_i^{\gamma}}\,^T({\bf r})$ does not impose any  conditions on 
$\varphi_C$. On the basis of these considerations, we can write the resolvent field as 
\begin{equation}
\overline{{\cal{A}}_{i}^{\gamma}}({\bf{r}})={\delta}_{{i}\,{\gamma}}\,\left({\varphi}_{A}+\Phi\right)+
\frac{r_i\,r_{\gamma}}{r^2}\,\left({\varphi}_{B}+r\Phi^{\prime}\right)+
{\epsilon}_{i{\gamma}n}\frac{r_n}{r}\,{\varphi}_{C}\,,
\label{eq:nuxistruct}
\end{equation}
and $\overline{{\cal{Y}}^\alpha}({\bf{r}})$ becomes
\begin{equation}
\overline{{\cal{Y}}^\alpha}({\bf{r}})=r_{\alpha}\,{\Phi}(r)\,.
\label{eq:Yeq}
\end{equation}
The same analysis can be applied to the gauge field $A_i^{\gamma}({\bf{r}})$, with the result 
that we obtain the wholly analogous equations
 \begin{equation}
A_i^{\gamma}\,^L({\bf r})={\delta}_{{i}\,{\gamma}}\,{\cal S}(r)+
\frac{r_i\,r_{\gamma}}{r}\,{\cal S}^{\prime}(r)
\label{eq:longa}
\end{equation}
and 
\begin{equation}
A_i^{\gamma}\,^T({\bf r})={\delta}_{{i}\,{\gamma}}\,{\cal T}_{A}(r)+
\frac{r_i\,r_{\gamma}}{r^2}\,{\cal T}_{B}(r)+{\epsilon}_{i{\gamma}n}\frac{r_n}{r}\,{\cal T}_{C}(r)
\label{eq:transa}
\end{equation}
with 
\begin{equation}
\frac{d(r^2{\cal T}_B)}{dr}+r^2\frac{d\,{\cal T}_A}{dr}=0\,.
\label{eq:trans}
\end{equation}
The complete gauge field can be represented as 
\begin{equation}
{{A}_{i}^{\gamma}}({\bf{r}})=\frac{1}{gr}\left\{\left({\delta}_{{i}\,{\gamma}}-
\frac{r_ir_{\gamma}}{r^2}\right)\,
{\cal N}+\frac{r_i\,r_{\gamma}}{r}\frac{d\,{\cal N}}{dr}\right\}+
{\delta}_{{i}\,{\gamma}}\,{\cal T}_{A}+
\frac{r_i\,r_{\gamma}}{r^2}\,{\cal T}_{B}+
{\epsilon}_{i{\gamma}n}\frac{r_n}{r}\,{\cal T}_{C}\,,
\label{eq:Atotal}
\end{equation}
where we have used Eqs.(\ref{eq:defN}) and (\ref{eq:longa}) to set $gr{\cal S}={\cal N}$. 
\bigskip

We can use the representations given in Eqs.(\ref{eq:longA}), (\ref{eq:transA}), (\ref{eq:longa}), and 
(\ref{eq:transa}) to express Eq.(\ref{eq:a2Y}) as
\begin{eqnarray}
\overline{{\cal{A}}_{i}^{\gamma}}\,_{\overline{\cal Y}}&=&
\left\{{\epsilon}_{i{\gamma}n}
\frac{r_n}{r}\left[{\varphi}_A+{\cal T}_A\right]-\left({\delta}_{i\,{\gamma}}-\frac{r_i\,r_{\gamma}}
{r^2}\right)\left[{\varphi}_C+{\cal T}_C\right]\right\}
\sin{\overline{\cal{N}}}\nonumber \\
&+&\frac{1}{gr}\,{\epsilon}_{i{\gamma}n}\,\frac{r_n}{r}\,\left(1-\cos\overline{\cal{N}}\right)\nonumber \\
&+&\left\{\left({\delta}_{i\,{\gamma}}-\frac{r_i\,r_{\gamma}}{r^2}\right)
\left[{\varphi}_A+{\cal T}_A\right]+{\epsilon}_{i{\gamma}n}\frac{r_n}{r}
\left[{\varphi}_C+{\cal T}_C\right]\right\}\left(1-\cos\overline{\cal{N}}\right)\nonumber \\
&+&\frac{1}{gr}\left({\delta}_{i\,{\gamma}}-\frac{r_i\,r_{\gamma}}{r^2}\right)\left(\overline{\cal{N}}
-\sin\overline{\cal{N}}\right)\,,
\label{eq:aypart}
\end{eqnarray}  
where we have used Eqs.(\ref{eq:defNbar}) and (\ref{eq:Yeq}) to set $gr{\Phi}=\overline{\cal{N}}$.
In a similar way, we can express Eq.(\ref{eq:a2X}) as
\begin{eqnarray}
\overline{{\cal{A}}_{i}^{\gamma}}\,_{\overline{\cal X}}&=&
\left\{{\epsilon}_{i{\gamma}n}
\frac{r_n}{r}\,\left[\frac{{\cal N}}{gr}+{\cal T}_A\right]-
\left({\delta}_{i\,{\gamma}}-\frac{r_i\,r_{\gamma}}
{r^2}\right){\cal T}_C\right\}
\sin{{\cal{N}}}\nonumber \\
&-&\frac{1}{gr}\,{\epsilon}_{i{\gamma}n}\,\frac{r_n}{r}\,\left(1-\cos{\cal{N}}\right)\nonumber \\
&-&\left\{\left({\delta}_{i\,{\gamma}}-\frac{r_i\,r_{\gamma}}{r^2}\right)
\left[\frac{{\cal N}}{gr}+{\cal T}_A\right]+{\epsilon}_{i{\gamma}n}\frac{r_n}{r}
{\cal T}_C\right\}\left(1-\cos{\cal{N}}\right)\nonumber \\
&+&\frac{1}{gr}\left({\delta}_{i\,{\gamma}}-\frac{r_i\,r_{\gamma}}{r^2}\right)\left({\cal{N}}
-\sin{\cal{N}}\right)\,.
\label{eq:axpart}
\end{eqnarray} 
It is manifest from Eqs.(\ref{eq:aypart}) and (\ref{eq:axpart}) that  
$r_{\gamma}\overline{{\cal{A}}_{i}^{\gamma}}({\bf{r}})_{\overline{\cal X}}=0$ and 
$r_{\gamma}\overline{{\cal{A}}_{i}^{\gamma}}({\bf{r}})_{\overline{\cal Y}}=0$, so that 
$r_{\gamma}\overline{{\cal{A}}_{i}^{\gamma}}({\bf{r}})=0$ too. When we use that fact in 
Eq.(\ref{eq:nuxistruct}), we find that 
\begin{equation}
{\Phi}+r{\Phi}^{\prime}+{\varphi}_A+{\varphi}_B=0\;\;\;\;\mbox{or, equivalently,}\;\;\;\;
\frac{\overline{\cal{N}}^{\,\prime}}{g}+{\varphi}_A+{\varphi}_B=0\,,
\label{eq:phib}
\end{equation}
enabling us to represent $\overline{{\cal{A}}_{i}^{\gamma}}({\bf{r}})$
as
\begin{equation}
\overline{{\cal{A}}_{i}^{\gamma}}({\bf{r}})=\left({\delta}_{i\,{\gamma}}-\frac{r_i\,r_{\gamma}}{r^2}\right)
\left(\frac{\overline{\cal{N}}}{gr}+{\varphi}_A\right)+{\epsilon}_{i{\gamma}n}\frac{r_n}{r}\,{\varphi}_C\,.
\label{eq:aform}
\end{equation}
Substitution of Eqs.(\ref{eq:aypart}), (\ref{eq:axpart}), and (\ref{eq:aform}) into Eq.(\ref{eq:azero}) 
enables us to express ${\varphi}_A$  and ${\varphi}_C$ as functions of $\overline{\cal{N}}$, ${\cal{N}}$, 
${\cal T}_A$, and 
${\cal T}_C$, so that we obtain 
\begin{equation}
\varphi_A=\frac{1}{gr}\left[{\cal{N}}{\cos}(\overline{\cal{N}}+{\cal{N}})-{\sin}(\overline{\cal{N}}+{\cal{N}})\right]
+{\cal T}_A\left[{\cos}(\overline{\cal{N}}+{\cal{N}})-1\right]-{\cal T}_C\,{\sin}(\overline{\cal{N}}+{\cal{N}})
\label{eq:phia}
\end{equation} 
and
\begin{equation}
\varphi_C=\frac{1}{gr}\left[{\cal{N}}{\sin}(\overline{\cal{N}}+{\cal{N}})+{\cos}(\overline{\cal{N}}+{\cal{N}})-1\right]
+{\cal T}_C\left[{\cos}(\overline{\cal{N}}+{\cal{N}})-1\right]+{\cal T}_A\,{\sin}(\overline{\cal{N}}+{\cal{N}})\,.
\label{eq:phic}
\end{equation}
We can solve Eqs.(\ref{eq:transcond}) and (\ref{eq:phib}) simultaneously to eliminate ${\varphi}_B$, and obtain 
a nonlinear differential equation that is equivalent to the nonlinear integral equation described in Eqs.(\ref{eq:azero})-(\ref{eq:a2Y})
{\em modulo} the previously-mentioned assumptions about the forms of the resolvent field and the gauge field. 
It is most convenient to use the dimensionless variable $u={\ln}(r/r_0)$, where $r_0$ is an arbitrary 
constant length, as the independent variable in expressing this differential equation, which then is 
\begin{eqnarray}
\frac{d^2\,\overline{\cal{N}}}{du^2}&+&\frac{d\,\overline{\cal{N}}}{du}+
2\left[{\cal{N}}{\cos}(\overline{\cal{N}}+{\cal{N}})-
{\sin}(\overline{\cal{N}}+{\cal{N}})\right]\nonumber \\
&+&2gr_0\exp(u)\left\{{\cal T}_A\left[{\cos}(\overline{\cal{N}}+{\cal{N}})-1\right]-{\cal T}_C\,
{\sin}(\overline{\cal{N}}+{\cal{N}})\right\}=0.
\label{eq:nueq}
\end{eqnarray}
Eq.(\ref{eq:nueq}) is a nonlinear differential equation in the variable $\overline{\cal{N}}$, which 
determines the resolvent field. ${\cal{N}}$, ${\cal T}_A$, and ${\cal T}_C$ will be taken to be known 
functions in this equation. One question of immediate interest is whether
there are nonvanishing solutions for the case that ${\cal{N}}$, ${\cal T}_A$, and ${\cal T}_C$ all vanish,
so that the gauge field  $A_i^{\gamma}$ is identically zero. In 
that case, Eq.(\ref{eq:nueq}) reduces to 
\begin{equation}
\frac{d^2\,\overline{\cal{N}}}{du^2}+\frac{d\,\overline{\cal{N}}}{du}-
2{\sin}(\overline{\cal{N}})=0.
\label{eq:nueqzero}
\end{equation}
We observe that Eq.(\ref{eq:nueqzero}) is the equation for a 
damped pendulum in which small-amplitude 
oscillations have not been assumed, in which $u$ corresponds 
to the time,  and $\overline{\cal{N}}=\theta+\pi$,
where $\theta=0$ when the 
pendulum is in its stable equilibrium position. Moreover, Eq.(\ref{eq:nueqzero}) corresponds to the 
equation for the pendulum in which mass, length, local acceleration due to gravity, and damping constant 
have been set $=1$, so that no adjustable parameters remain in the equation.  \bigskip 

In discussing Eqs.(\ref{eq:nueq}) and (\ref{eq:nueqzero}), we will assume that ${\cal T}_A$,
${\cal T}_C$, and ${\cal N}-2{\ell}\pi$ (where $\ell$ is an integer), 
vanish faster than $1/r$ as $r{\rightarrow}\infty$. Moreover, we also 
assume that the gauge field is bounded everywhere, so that 
$r{\cal T}_A{\rightarrow}0$ as $r{\rightarrow}0$, and that the same applies to $r{\cal T}_C$, and
$r{\cal N}$. We will make these conditions more precise in the Appendix. \bigskip

The fact that Eq.(\ref{eq:nueqzero}) describes a damped pendulum moving without a driving force 
makes it obvious that it has nonvanishing solutions. In the limit $u{\rightarrow}\infty$, 
the damped pendulum must come to rest in a position of  static equilibrium, so 
that, as $u{\rightarrow}\infty$, $\overline{\cal{N}}{\rightarrow}m\pi$, where $m$ is an arbitrary 
integer. Physical considerations lead us to expect that the damped pendulum will 
come to rest in a position of stable equilibrium, 
for which $\overline{\cal{N}}{\rightarrow}(2m+1)\pi$. But in the application of Eq.(\ref{eq:nueqzero})
of interest to us in the current context, all possible solutions --- those that 
terminate at saddle points as well as those that terminate at stable
equilibrium positions --- must be considered as long as they 
are bounded functions in the interval $-\infty<u<\infty$. Eq. (\ref{eq:nueq}) is more complicated than 
Eq.(\ref{eq:nueqzero}); but as a result of the conditions on ${\cal T}_A$, and ${\cal T}_C$,
and on ${\cal N}$ for the $\ell=0$ case, 
discussed above and made precise in the Appendix,  
Eq. (\ref{eq:nueq}) approaches Eq.(\ref{eq:nueqzero}) at the 
origin in configuration space, $i.\,e.$ in the limit $u{\rightarrow}-\infty$, as well as 
in the limit $u{\rightarrow}\infty$. It 
is therefore reasonable to expect that $\overline{\cal{N}}{\rightarrow}m\pi$ as $u{\rightarrow}\infty$
even when it is governed by Eq.(\ref{eq:nueq}). 
That this expectation is fulfilled, is shown in the Appendix.  When $\ell{\neq}0$, 
$\overline{\cal{N}}$ still approaches a definite limit as $u{\rightarrow}\infty$, but that limit is 
no longer an integer multiple of $\pi$. This case is discussed in Section \ref{sec-top}, and 
the mathematical foundations for this discussion are given in the Appendix. 
\bigskip

It is possible to use the results obtained in this section, in particular Eqs.(\ref{eq:transA}), (\ref{eq:transa}), 
(\ref{eq:phia}) and (\ref{eq:phic}), to represent the gauge-invariant gauge field as 
\begin{eqnarray}
&&A_{{\sf GI}\,i}^{\gamma}({\bf{r}})=\frac{1}{gr}\left\{\epsilon_{i\,\gamma\,n}\frac{r_n}{r}
\left[{\cos}(\overline{\cal{N}}+{\cal{N}})-1+{\cal{N}}\sin(\overline{\cal{N}}+{\cal{N}})\right]
+\left({\delta}_{i\,\gamma}-\frac{r_ir_{\gamma}}{r^2}\right)\times\right.\nonumber\\
&&\left.\times\left[{\cal{N}}{\cos}(\overline{\cal{N}}+{\cal{N}})-{\sin}(\overline{\cal{N}}+{\cal{N}})\right]
-\frac{r_ir_{\gamma}}{r}\frac{d\overline{\cal{N}}}{dr}\right\}\nonumber \\
&&+{\cal T}_A\left\{
\left({\delta}_{i\,\gamma}-\frac{r_ir_{\gamma}}{r^2}\right)
{\cos}(\overline{\cal{N}}+{\cal{N}})
+\epsilon_{i\,\gamma\,n}\frac{r_n}{r}{\sin}(\overline{\cal{N}}+{\cal{N}})\right\}+\frac{r_ir_{\gamma}}{r^2}
\left({\cal T}_A+{\cal T}_B\right)\nonumber \\
&&+{\cal T}_C\left\{\epsilon_{i\,\gamma\,n}\frac{r_n}{r}{\cos}(\overline{\cal{N}}+{\cal{N}})
-\left({\delta}_{i\,\gamma}-\frac{r_ir_{\gamma}}{r^2}\right)
{\sin}(\overline{\cal{N}}+{\cal{N}})\right\}\,.
\label{eq:AGIsub}
\end{eqnarray}
When we set the entire gauge field $A_i^{\gamma}=0$, we obtain
\begin{equation}
\left[A_{{\sf GI}\,i}^{\gamma}({\bf{r}})\right]_{(A_i^{\gamma}=0)}=\frac{1}{gr}\left\{\epsilon_{i\,\gamma\,n}\frac{r_n}{r}
\left[{\cos}(\,\overline{\cal{N}}\,)-1\right]-\left({\delta}_{i\,\gamma}
-\frac{r_ir_{\gamma}}{r^2}\right){\sin}\overline{\cal{N}}-\frac{r_ir_{\gamma}}{r}
\frac{d\overline{\cal{N}}}{dr}\right\}\,.
\label{eq:AGInoA}
\end{equation}
As a consequence of Eq.(\ref{eq:nueq}), ${\partial}_iA_{{\sf GI}\,i}^{\gamma}({\bf{r}})=0$, and
similarly, ${\partial}_i\left[A_{{\sf GI}\,i}^{\gamma}({\bf{r}})\right]_{(A_i^{\gamma}=0)}=0$ 
follows from Eq.(\ref{eq:nueqzero}),  
so that the 
consistency of Eqs.(\ref{eq:AGIsub}) and (\ref{eq:AGInoA}) with the 
transversality of the gauge-invariant gauge field is confirmed. \bigskip 

In order to investigate the changes in $\overline{\cal{N}}$ that can ensue from gauge transformations, 
we turn our attention to  the changes 
${\delta}{\cal{N}}$, ${\delta}{\cal T}_A$, ${\delta}{\cal T}_B$, 
and  ${\delta}{\cal T}_C$ in the corresponding quantities ${\cal{N}}$, ${\cal T}_A$, ${\cal T}_B$, 
and ${\cal T}_C$ due to the infinitesimal SU(2) gauge transformation
\begin{equation}
{\delta}A_i^{\gamma}({\bf r})={\partial}_i{\delta}{\omega}^{\gamma}({\bf r})+
g{\epsilon}_{{\gamma}{\alpha}{\beta}}
A_i^{\alpha}({\bf r}){\delta}{\omega}^{\beta}({\bf r}).
\label{eq:gaugetrans}
\end{equation}
We extend the {\em ansatz} that led to Eqs.(\ref{eq:longa}) and (\ref{eq:transa})
to choose an expression for ${\delta}{\omega}^{\gamma}({\bf r})$ that is isotropic except in so far 
as it transforms as a vector in the adjoint representation of SU(2). We therefore 
represent ${\delta}{\omega}^{\gamma}({\bf r})$ as 
\begin{equation}
{\delta}{\omega}^{\gamma}({\bf r})=\frac{r_{\gamma}}{r}{\Delta}( r),
\label{eq:delomega}
\end{equation}
where ${\Delta}( r)$ is an infinitesimal, isotropic quantity.
We substitute ${\delta}{\cal{N}}$, ${\delta}{\cal T}_A$, ${\delta}{\cal T}_B$, 
and  ${\delta}{\cal T}_C$ for ${\cal{N}}$, ${\cal T}_A$, ${\cal T}_B$, 
and ${\cal T}_C$ respectively, in Eq.(\ref{eq:Atotal}), to represent 
${\delta}A_i^{\gamma}({\bf r})$, the infinitesimal change in $A_i^{\gamma}({\bf r})$ due to a
gauge transformation, and obtain
\begin{equation}
{\delta}A_i^{\gamma}({\bf r})=\left({\delta}_{i\,{\gamma}}-\frac{r_i\,r_{\gamma}}{r^2}\right)
\frac{{\delta}{\cal N}}{gr}+\frac{r_i\,r_{\gamma}}{gr^2}\frac{d\,({\delta}{\cal N})}{dr}
+{\delta}_{{i}\,{\gamma}}\,{\delta}{\cal T}_{A}(r)+
\frac{r_i\,r_{\gamma}}{r^2}\,{\delta}{\cal T}_{B}(r)+
{\epsilon}_{i{\gamma}n}\frac{r_n}{r}\,{\delta}{\cal T}_{C}(r)\,.
\label{eq:deltaAi}
\end{equation}
 We then substitute Eq.(\ref{eq:deltaAi}) in the left-hand side of Eq. (\ref{eq:gaugetrans}),
and  Eq.(\ref{eq:Atotal}) in the right-hand side to obtain
\begin{equation}
\frac{{\delta}{\cal{N}}(r)}{gr}+{\delta}{\cal T}_A(r)=\frac{{\Delta}(r)}{r}+g{\Delta}(r){\cal T}_C(r),
\label{eq:delTA}
\end{equation}
\begin{equation}
r\frac{d}{dr}\left(\frac{{\delta}{\cal{N}}(r)}{gr}\right)+{\delta}{\cal T}_B(r)=-\frac{{\Delta}(r)}{r}+
\frac{d}{dr}{{\Delta}}(r)-g{\Delta}(r){\cal T}_C(r),
\label{eq:delTB}
\end{equation}
and
\begin{equation}
{\delta}{\cal T}_C(r)=-g{\Delta}(r)\left(\frac{{\cal{N}}(r)}{gr}+{\cal T}_A(r)\right).
\label{eq:delTC}
\end{equation}
We combine Eqs.(\ref{eq:trans}), (\ref{eq:delTA}) and (\ref{eq:delTB}) to obtain
\begin{equation}
r^2\frac{d^2}{dr^2}{\delta}{\phi}(r)+2r\frac{d}{dr}{\delta}{\phi}(r)-2{\delta}{\phi}(r)
=-2gr{\Delta}(r)T_C(r),
\label{eq:phieq}
\end{equation}       
where we have set 
\begin{equation}
\frac{{\delta}{\cal{N}}(r)}{g}-{\Delta}(r)={\delta}{\phi}(r).
\label{eq:NDphi}
\end{equation}
With a little more algebra we can also obtain
\begin{equation}
{\delta}{\cal T}_A(r)=-\frac{r}{2}\frac{d^2}{dr^2}{\delta}{\phi}(r)-\frac{d}{dr}{\delta}{\phi}(r)
\label{eq:delTAeq}
\end{equation}
and 
\begin{equation}
{\delta}{\cal T}_B(r)=\frac{r}{2}\frac{d^2}{dr^2}{\delta}{\phi}(r).
\label{eq:delTBeq}
\end{equation}
To investigate the changes that gauge transformations induce in $\overline{\cal{N}}$, we 
examine Eqs.(\ref{eq:phia}) and (\ref{eq:phic}) to find relationships between the changes that 
infinitesimal gauge transformations produce in the quantities that appear in these two equations. 
We find that
\begin{eqnarray}
{\delta}\varphi_A&&=\frac{1}{gr}\left\{{\delta}{\cal{N}}{\cos}(\overline{\cal{N}}+{\cal{N}})
-\left[{\cal{N}}{\sin}(\overline{\cal{N}}+{\cal{N}})+{\cos}(\overline{\cal{N}}+{\cal{N}})\right]
({\delta}\overline{\cal{N}}+{\delta}{\cal{N}})\right\}\nonumber\\
&&+\,{\delta}{\cal T}_A\left[{\cos}(\overline{\cal{N}}+{\cal{N}})-1\right]-
{\cal T}_A\,{\sin}(\overline{\cal{N}}+{\cal{N}})({\delta}\overline{\cal{N}}+{\delta}{\cal{N}})\nonumber\\
&&-\,{\delta}{\cal T}_C\,{\sin}(\overline{\cal{N}}+{\cal{N}})-{\cal T}_C\,{\cos}(\overline{\cal{N}}+
{\cal{N}})({\delta}\overline{\cal{N}}+{\delta}{\cal{N}})
\label{eq:deltaphia}
\end{eqnarray} 
and 
\begin{eqnarray}
{\delta}\varphi_C&&=\frac{1}{gr}\left\{{\delta}{\cal{N}}{\sin}(\overline{\cal{N}}+{\cal{N}})
+\left[{\cal{N}}{\cos}(\overline{\cal{N}}+{\cal{N}})-{\sin}(\overline{\cal{N}}+{\cal{N}})\right]
({\delta}\overline{\cal{N}}+{\delta}{\cal{N}})\right\}\nonumber\\
&&+\,{\delta}{\cal T}_C\left[{\cos}(\overline{\cal{N}}+{\cal{N}})-1\right]-
{\cal T}_C\,{\sin}(\overline{\cal{N}}+{\cal{N}})({\delta}\overline{\cal{N}}+{\delta}{\cal{N}})\nonumber\\
&&+\,{\delta}{\cal T}_A\,{\sin}(\overline{\cal{N}}+{\cal{N}})+{\cal T}_A\,{\cos}(\overline{\cal{N}}+
{\cal{N}})({\delta}\overline{\cal{N}}+{\delta}{\cal{N}})\,.
\label{eq:deltaphic}  
\end{eqnarray}
As shown in Eq.(\ref{eq:GIglue}), changes produced by gauge transformations 
in the  transverse components of $A_i^{\gamma}$ and 
$\overline{\cal A}_i^{\gamma}$ cancel, so that
\begin{equation}
{\delta}\varphi_A+{\delta}{\cal T}_A=0,\;\;\;{\delta}\varphi_B+{\delta}{\cal T}_B=0,\;\;\mbox{and}\;
\;{\delta}\varphi_C+{\delta}{\cal T}_C=0.
\label{eq:deltasums}
\end{equation}
We substitute Eq.(\ref{eq:deltasums}) in Eqs.(\ref{eq:deltaphia}) and (\ref{eq:deltaphic}),
set
\begin{equation}
X={\cal T}_C+\frac{1}{gr},\;\;Y={\cal T}_A+\frac{{\cal{N}}}{gr},\;\;
{\theta}=\overline{\cal{N}}+{\cal{N}},\;\;\mbox{and}\;\;
{\xi}={\delta}\overline{\cal{N}}+{\delta}{\cal{N}}-g{\Delta}\,,
\label{eq:subst}
\end{equation}
and find that
\begin{equation}
\left(X{\cos}{\theta}+Y{\sin}{\theta}\right){\xi}=0
\label{eq:subphia}
\end{equation}
and
\begin{equation}
\left(X{\sin}{\theta}-Y{\cos}{\theta}\right){\xi}=0\,.
\label{eq:subphic}
\end{equation} 
Since $X$ and $Y$ are determined by the initial configuration of the gauge field $A_i^{\gamma}$,
which is arbitrary, Eqs.(\ref{eq:subphia}) and (\ref{eq:subphic}) must be satisfied without
requiring that $X$ and $Y$ vanish. It therefore follows that $\xi=0$, which, with Eq.(\ref{eq:NDphi}), leads to 
\begin{equation}
{\delta}\overline{\cal{N}}=-g{\delta}{\phi},
\label{eq:Nphi}
\end{equation}
and, with Eq.(\ref{eq:phieq}), to the following differential equation 
for the variation of $\overline{\cal{N}}$
produced by an infinitesimal gauge transformation:
\begin{equation}
r^2\frac{d^2}{dr^2}{\delta}\overline{\cal{N}}(r)+2r\frac{d}{dr}{\delta}\overline{\cal{N}}(r)-
2{\delta}\overline{\cal{N}}(r)=2g^2{\Delta}(r)r{\cal T}_C(r).
\label{eq:Nbareq}
\end{equation}
Further confirmation of the consistency of these results is obtained by examining the variation
of $A_{{\sf GI}\,i}^{\gamma}({\bf{r}})$
due to an infinitesimal gauge transformation of 
the quantities that constitute it. This variation is given by 
\begin{eqnarray}
&&{\delta}A_{{\sf GI}\,i}^{\gamma}({\bf{r}})=\frac{1}{gr}\left\{\epsilon_{i\,\gamma\,n}\frac{r_n}{r}
\left[{\delta}{\cal{N}}\sin(\overline{\cal{N}}+{\cal{N}})+\left({\cal{N}}{\cos}\,(\overline{\cal{N}}+
{\cal{N}})-\sin(\overline{\cal{N}}+{\cal{N}})\right)
{\delta}\left(\overline{\cal{N}}+{\cal{N}}\right)\right]+\right.\nonumber\\
&&\left.\left({\delta}_{i\,\gamma}-\frac{r_ir_{\gamma}}{r^2}\right)
\left[{\delta}{\cal{N}}{\cos}(\overline{\cal{N}}+{\cal{N}})-\left({\cal{N}}{\sin}\,(\overline{\cal{N}}+
{\cal{N}})+\cos(\overline{\cal{N}}+{\cal{N}})\right)
{\delta}\left(\overline{\cal{N}}+{\cal{N}}\right)\right]
-\frac{r_ir_{\gamma}}{r}\frac{d({\delta}\overline{\cal{N}})}{dr}\right\}\nonumber \\
&&+{\delta}{\cal T}_A\left\{
\left({\delta}_{i\,\gamma}-\frac{r_ir_{\gamma}}{r^2}\right)
{\cos}(\overline{\cal{N}}+{\cal{N}})
+\epsilon_{i\,\gamma\,n}\frac{r_n}{r}{\sin}(\overline{\cal{N}}+
{\cal{N}})\right\}+\frac{r_ir_{\gamma}}{r^2}
\left({\delta}{\cal T}_A+{\delta}{\cal T}_B\right)\nonumber \\
&&+{\delta}{\cal T}_C\left\{\epsilon_{i\,\gamma\,n}\frac{r_n}{r}{\cos}(\overline{\cal{N}}+{\cal{N}})
-\left({\delta}_{i\,\gamma}-\frac{r_ir_{\gamma}}{r^2}\right){\sin}(\overline{\cal{N}}+
{\cal{N}})\right\}\nonumber\\
&&-{\cal T}_A\left\{
\left({\delta}_{i\,\gamma}-\frac{r_ir_{\gamma}}{r^2}\right)
{\sin}(\overline{\cal{N}}+{\cal{N}})
-\epsilon_{i\,\gamma\,n}\frac{r_n}{r}{\cos}(\overline{\cal{N}}+{\cal{N}})\right\}
{\delta}\left(\overline{\cal{N}}+{\cal{N}}\right)\nonumber \\
&&-{\cal T}_C\left\{\epsilon_{i\,\gamma\,n}\frac{r_n}{r}{\sin}(\overline{\cal{N}}+{\cal{N}})
+\left({\delta}_{i\,\gamma}-\frac{r_ir_{\gamma}}{r^2}\right)
{\cos}(\overline{\cal{N}}+{\cal{N}})\right\}{\delta}\left(\overline{\cal{N}}+{\cal{N}}\right)\,.
\label{eq:AGIdelta}
\end{eqnarray}
When  Eqs.(\ref{eq:phieq}), (\ref{eq:NDphi}), (\ref{eq:delTAeq}), (\ref{eq:delTBeq}), and 
(\ref{eq:Nbareq}) are used  to replace ${\delta}{\cal{N}}$, ${\delta}\overline{\cal{N}}$, 
${\delta}{\cal T}_A$, ${\delta}{\cal T}_B$, and ${\delta}{\cal T}_C$ in Eq.(\ref{eq:AGIdelta}), 
we find that   
${\delta}A_{{\sf GI}\,i}^{\gamma}({\bf{r}})=0$, confirming the consistency of the procedure that 
leads to Eq.(\ref{eq:Nbareq}). \bigskip

The effect of small gauge transformations on the limits of ${\delta}\overline{\cal{N}}$ as 
$u{\rightarrow}\pm\infty$ can best be explored by expressing Eq.(\ref{eq:Nbareq}) in terms of 
the independent variable $u$. As is discussed in the Appendix, the expression $r{\cal T}_C$ appearing 
on the right-hand-side of that equation is subject to the conditions          
$\lim_{u{\rightarrow}\infty}|r_0\exp(u){\cal T}_C(u)|<K\exp(-{\alpha}u)$ and             
$\lim_{u{\rightarrow}-\infty}|r_0\exp(u){\cal T}_C(u)|<K^{\prime}\exp({\alpha}u)$, 
where $K$ and $K^{\prime}$
are constants. Constraints can be placed on $\Delta$, but the only one necessary to complete this 
argument is that $\Delta$ is bounded in the interval $(-\infty,\, \infty)$. These circumstances 
allow us to conclude that Eq.(\ref{eq:Nbareq}) can be represented, 
in the limits $u{\rightarrow}\infty$ and $u{\rightarrow}-\infty$,  in the form
\begin{equation}
\left|\frac{d^2}{du^2}{\delta}\overline{\cal{N}}(u)+\frac{d}{du}{\delta}\overline{\cal{N}}(u)-
2{\delta}\overline{\cal{N}}(u)\right|\,{\leq}\,C\exp(-{\alpha}|u|)\,,
\label{eq:Nbaru}
\end{equation}
where $C$ is a constant.  We can therefore infer that 
\begin{equation} 
\lim_{u{\rightarrow}-\infty}|{\delta}\overline{\cal{N}}(u)|\,{\leq}\,A\exp({\alpha}u)+B\exp(u)
\;\;\mbox{and}\:\: \lim_{u{\rightarrow}\infty}|{\delta}\overline{\cal{N}}(u)|\,{\leq}\,
A^{\prime}\exp({-\alpha}u)+B^{\prime}{\exp}(-2u)
\label{eq:Nbaruinf}
\end{equation}
for some constants $A$, $B$, $A^{\prime}$, and $B^{\prime}$. 
We observe that the limiting values of 
${\delta}\overline{\cal{N}}(u)$, as $u{\rightarrow}{\pm}\infty$, vanish, 
and that therefore the corresponding limiting 
values of $\overline{\cal{N}}(u)$, as $u{\rightarrow}\infty$ 
and as $u{\rightarrow}-\infty$, are invariant to small gauge transformations --- $i.\,e.$
transformations implemented by $\partial_i{\Pi}^{\gamma}_i({\bf{r}})+g{\epsilon}_{\gamma\beta\alpha}
A_i^{\beta}({\bf{r}}){\Pi}_i^{\alpha}({\bf{r}})+g{\psi}^{\dagger}({\bf{r}})
\frac{{\tau}^{\gamma}}{2}{\psi}({\bf{r}})$ acting as the generator of infinitesimal 
gauge transformations. But $\overline{\cal{N}}(u)$, for arbitrary values of $u$, 
does not share that invariance.  
  These arguments will be used, in the next section, to 
show that the winding number of the gauge-invariant gauge field 
remains unaffected by  small gauge transformations.

\section{topology and the implementation of gauge invariance}
\label{sec-top}
Our considerations in this section are based on the fact, discussed in Section~\ref{sec-config}, that 
the transformation that converts the quark field ${\psi}({\bf{r}})$ and the gauge field
$A_i^{\gamma}({\bf{r}})$ into their respective gauge-invariant forms --- as shown in 
Eqs.(\ref{eq:GIpsi}) and (\ref{eq:AdressedAxz}) respectively --- has the formal 
structure of a gauge transformation, although it
differs from a gauge transformation in a very important 
respect: The argument of the operator-valued
$V_{\cal{C}}({\bf{r}})$ that effects the transformation to gauge-invariant forms 
is not a function by which ${\psi}({\bf{r}})$ and $A_{i}^{\gamma}({\bf{r}})$
are gauge-transformed, but is itself operator-valued and subject to gauge transformations.
Eqs.(\ref{eq:GIpsi}) and (\ref{eq:AdressedAxz}) therefore actually represent transformations of 
gauge-dependent fields to forms that are 
invariant to further gauge transformations. Nevertheless, 
when we represent $V_{\cal{C}}({\bf{r}})$
as a number-valued realization of the operator-valued quantity, we can use its formal structure to 
investigate the topological features of the gauge-invariant gauge fields. 
\bigskip

 With the {\em ansatz} that led to the expression for the resolvent field given in Eq. (\ref{eq:aform}),
we find that $V_C({\bf{r}})$ is given by
\begin{equation}
V_C({\bf{r}})={\exp}\left(-i{\hat r}_n{\tau}_n\frac{\left(\overline{\cal{N}}+
{\cal{N}}\right)}{2}\right)\,.
\label{eq:VCY}
\end{equation}
and that the expression for $A_{{\sf GI}\,i}^{\gamma}({\bf{r}})$ given in Eq.(\ref{eq:AGIsub}) 
can be written as 
\begin{equation}
-ig\frac{\tau^{\gamma}}{2}A_{{\sf GI}\,i}^{\gamma}({\bf{r}})=-igV_C({\bf{r}})\frac{\tau^{\gamma}}{2}
A_{i}^{\gamma}({\bf{r}})V_C^{-1}({\bf{r}})+V_C({\bf{r}}){\partial}_i
V_C^{-1}({\bf{r}}),
\label{eq:AGIVform}
\end{equation}
where $A_{i}^{\gamma}({\bf{r}})$ is given in Eqs.(\ref{eq:Atotal}), and where we can define 
\begin{equation}
{\sf A}\,_i({\bf{r}})=-ig\frac{{\tau}^{\gamma}}{2}\left[A_{{\sf GI}\,i}^{\gamma}({\bf{r}})\right]_{V}
=V_C({\bf{r}}){\partial}_iV^{-1}_C({\bf{r}})
\label{eq:Ufieldformal}
\end{equation}
so that  $\left[A_{{\sf GI}\,i}^{\gamma}({\bf{r}})\right]_{V}$ is the part of 
$A_{{\sf GI}\,i}^{\gamma}({\bf{r}})$ given by
\begin{equation}
\left[A_{{\sf GI}\,i}^{\gamma}({\bf{r}})\right]_{V}=\frac{1}{gr}\left\{\epsilon_{i\,\gamma\,n}
\frac{r_n}{r}
\left({\cos}(\,\overline{\cal{N}}+{\cal{N}}\,)-1\right)-\left({\delta}_{i\,\gamma}
-\frac{r_ir_{\gamma}}{r^2}\right){\sin}(\overline{\cal{N}}+{\cal{N}})-\frac{r_ir_{\gamma}}{r}\,
\frac{d(\overline{\cal{N}}+{\cal{N}})}{dr}\right\}\,.
\label{eq:Ufield}
\end{equation}
Eq.(\ref{eq:Ufieldformal}) shows that $\left[A_{{\sf GI}\,i}^{\gamma}({\bf{r}})\right]_{V}$ has the formal 
structure of the ``pure gauge'' part of a gauge field, although that is not what it is. 
$\left[A_{{\sf GI}\,i}^{\gamma}({\bf{r}})\right]_{V}$
is a part of the gauge-invariant field
$A_{{\sf GI}\,i}^{\gamma}({\bf{r}})$, and any change in $\left[A_{{\sf GI}\,i}^{\gamma}({\bf{r}})\right]_{V}$
due to an infinitesimal gauge transformation must be offset by compensating gauge transformations to
$V_C({\bf{r}})\frac{\tau^{\gamma}}{2}
A_{i}^{\gamma}({\bf{r}})V_C^{-1}({\bf{r}})$. We can, nevertheless, make
 use of the formal structure 
of ${\sf A}\,_i({\bf{r}})$ represented in Eq.(\ref{eq:Ufieldformal}) 
to define
\begin{equation} 
Q=-(24{\pi}^2)^{-1}{\epsilon}_{ijk}{\int}d{\bf{r}}\mbox{Tr}[{\sf A}\,_i({\bf{r}}){\sf A}\,_j({\bf{r}})
{\sf A}\,_k({\bf{r}})]\,;
\label{eq:Qformal}
\end{equation}
and, following Ref.\cite{Jack}, we can use Eqs.(\ref{eq:VCY}) and (\ref{eq:Ufieldformal}) to express $Q$ as 
\begin{equation}
Q=\frac{1}{4{\pi}^2}{\int}\frac{d{\bf{r}}}{r^2}{\sin}^2\left(\frac{\overline{\cal{N}}+{\cal{N}}}{2}\right)
\frac{d\,\left(\overline{\cal{N}}+{\cal{N}}\right)}{dr}\,,
\label{eq:Qnumber}
\end{equation}
and integrate this expression to obtain 
\begin{eqnarray}
Q=&&\frac{1}{2\pi}\lim_{r{\rightarrow}{\infty}}\left\{\overline{\cal{N}}(r)+{\cal{N}}(r)-
{\sin}\left(\overline{\cal{N}}(r)+{\cal{N}}(r)\right)
\right\}\nonumber\\
&&-\frac{1}{2\pi}\left\{\overline{\cal{N}}(0)+{\cal{N}}(0)-{\sin}
\left(\overline{\cal{N}}(0)+{\cal{N}}(0)\right)\right\}\,.
\label{eq:windingNbarN}
\end{eqnarray}
We will refer to $Q$ as the winding number of the gauge-invariant gauge field, even though its
values are not restricted to integers. Unlike the winding numbers of the gauge field, which are
integer-valued,\cite{Jack,Rho,Bhad,Zahed} $Q$ is determined by the properties 
of $\overline{\cal N}$, which is governed by Eq.(\ref{eq:nueq}) and by the fact that 
it must be bounded in the 
entire interval $0{\leq}r<\infty$. In discussing the properties of ${\cal N}$, we define the ``pure
gauge''  gauge field
\begin{equation}
A_{i}^{\gamma}({\bf{r}})_{{\sf p}{\sf g}}=\frac{1}{gr}\left\{\epsilon_{i\,\gamma\,n}\frac{r_n}{r}
\left[{\cos}(\,{\cal{N}}\,)-1\right]-\left({\delta}_{i\,\gamma}
-\frac{r_ir_{\gamma}}{r^2}\right){\sin}{\cal{N}}-\frac{r_ir_{\gamma}}{r}
\frac{d{\cal{N}}}{dr}\right\} 
\label{eq:Apg}
\end{equation}
which we obtain from
\begin{equation}
A_{i}^{\gamma}({\bf{r}})_{{\sf p}{\sf g}}=-\,\frac{i}{g}{\sf Tr}\left[{\tau}_{\gamma}\exp\left({-i{\cal N}
\frac{{\tau}_{\alpha}r_{\alpha}}{2r}}\right)\partial_i\exp\left({i{\cal N}
\frac{{\tau}_{\alpha}r_{\alpha}}{2r}}\right)\right].
\label{eq:Apgsource}
\end{equation} 
We further observe that, in Section \ref{sec-config} and in the Appendix, 
we have assumed that 
${\cal N}=0$ at $r=0$, and that $\lim_{\,r{\rightarrow}\infty}{\cal N}=2\ell\pi$, so that 
we consider gauge fields $A_{i}^{\gamma}({\bf{r}})_{{\sf p}{\sf g}}$ 
whose winding number is the integer $\ell$, and therefore that 
\begin{equation}
Q=\frac{1}{2\pi}\left\{\lim_{r{\rightarrow}{\infty}}\left[\overline{\cal{N}}(r)-
{\sin}\left(\overline{\cal{N}}(r)\,\right)\,\right]+2\pi\ell
-\overline{\cal{N}}(0)+{\sin}\left(\overline{\cal{N}}(0)\,\right)\right\}\,.
\label{eq:winding}
\end{equation}
As is shown in the Appendix, when $\ell=0$
and $\overline{\cal{N}}$ is normalized so that $\overline{\cal{N}}=0$ when $r=0$, 
$\lim_{\,r{\rightarrow}\infty}\overline{\cal{N}}=m\pi$. 
For the $\ell=0$ case, Q is therefore given by 
\begin{equation} 
Q=\frac{1}{2\pi}\left(\lim_{r{\rightarrow}{\infty}}\overline{\cal{N}}(r)-
\overline{\cal{N}}(0)\right)=\frac{1}{2\pi}\,\lim_{r{\rightarrow}{\infty}}\overline{\cal{N}}(r)
=\frac{m}{2}\,,   
\label{eq:wgnosin}
\end{equation}
 where $m$ is an integer. 
We will now discuss the implications of Eq.(\ref{eq:nueq}) for the values 
of $Q$ when $\ell{\neq}0$.
\bigskip

When we consider gauge fields with winding numbers $\ell{\neq}0$, 
for which $\lim_{u{\rightarrow}{\infty}}{\cal{N}}(u)=2\pi{\ell}$,
but $\lim_{u{\rightarrow}{\infty}}\exp(u){\cal T}_A(u)$, and 
$\lim_{u{\rightarrow}{\infty}}\exp(u){\cal T}_C(u)$ still vanish along with 
$\lim_{u{\rightarrow}{-\infty}}{\cal{N}}(u)$, 
$\lim_{u{\rightarrow}{-\infty}}\exp(u){\cal T}_A(u)$, and 
$\lim_{u{\rightarrow}{-\infty}}\exp(u){\cal T}_C(u)$, 
then we observe that, in the limit $u{\rightarrow}\infty$, 
 $\overline{\cal{N}}(u)$ satisfies 
\begin{equation}
\frac{d^2\,\overline{\cal{N}}}{du^2}+\frac{d\,\overline{\cal{N}}}{du}+
2\left[2{\pi}{\ell}{\cos}(\overline{\cal{N}}+2{\pi}{\ell})-
{\sin}(\overline{\cal{N}}+2{\pi}{\ell})\right]=0
\label{eq:Nasymp}
\end{equation}
and  $\overline{\cal{N}}(u)$ approaches a limiting value  
$\lim_{u{\rightarrow}{\infty}}\overline{\cal{N}}(u)$, 
which exists, but is not an integer multiple of $\pi$.
$\overline{\cal{N}}(u)$ vanishes, as before, at the saddle point for which $u{\rightarrow}-\infty$, 
but $\lim_{u{\rightarrow}{\infty}}\overline{\cal{N}}(u)$ now is one of the denumerably 
infinite solutions of the transcendental equation 
\begin{equation}
\tan\left\{\lim_{u{\rightarrow}{\infty}}\overline{\cal{N}}(u)\right\}=2\pi{\ell}\,.
\label{eq:transscend}
\end{equation}
If we adopt the convention that the inverse tangent is defined so that, for $\xi={\tan}^{-1}({\eta})$,
$-\frac{\pi}{2}\,<\,\xi\,<\,\frac{\pi}{2}$, then the winding numbers for the 
gauge-invariant gauge field are given by 
\begin{equation}
Q=\frac{1}{2\pi}\left\{{\tan}^{-1}(2{\pi}{\ell})+m\pi+2{\pi}{\ell}
\left(1-\frac{1}{\sqrt{1+4{\pi}^2{\ell}^2}}\right)\right\}
\label{eq:windnew}
\end{equation}
for some integer $m$.\bigskip

Eq.(\ref{eq:windnew}) establishes a winding number for the gauge-invariant gauge field 
that is a function of two integer-valued variables, $\ell$ and $m$, but that is not 
itself integer-valued.  The integer $\ell$ defines the homotopy class of the 
gauge field described by Eqs.(\ref{eq:Apg}) and (\ref{eq:Apgsource}). 
Gauge-invariant gauge fields can be categorized by
the value of $\ell$ that describes the homotopy class of the gauge fields to which they are 
linked by Eq.(\ref{eq:nueq}),
and by the value of $m$ that defines the sheet on which the inverse tangent function 
that corresponds to the $\lim_{u{\rightarrow}{\infty}}\overline{\cal{N}}(u)$ is located.
The range of values of the integer
$m$, for which $m\pi$ corresponds to limits that $\overline{\cal N}(u)$
can actually attain, depends on the solutions of Eq.(\ref{eq:nueq}) that 
given sets of functions ${\cal N}(u)$, ${\cal T}_A(u)$ and ${\cal T}_C(u)$ can support.
Examples of solutions, and their implications for winding numbers of the 
gauge-invariant gauge field, are given in Section \ref{sec-numres}.\bigskip

The relationship between 
$Q$, $\ell$ and $m$  reflects the fact that, 
when $\lim_{u{\rightarrow}-\infty}{\cal N}(u)=0$, the  
$\lim_{\,u{\rightarrow}\infty}{\cal N}(u)$ is 
the only part of the gauge field that can affect its homotopy class; however, the functional dependence 
of ${\cal N}(u)$, ${\cal T}_A(u)$ and ${\cal T}_C(u)$ for the entire range of values of $u$ affects the 
$\lim_{\,u{\rightarrow}\infty}\overline{\cal N}(u)$, and therefore $Q$, 
as shown in Eqs.(\ref{eq:nueq}) and 
(\ref{eq:winding}). A number of different values of $m$ can therefore be compatible with 
the same $\ell$, and, in fact, with the identical functions  
${\cal N}(u)$, ${\cal T}_A(u)$ and ${\cal T}_C(u)$.
The mathematical foundations for these observations are discussed in Section 
\ref{sec-numres} and in the Appendix.

\newcommand{\wN}{{\cal N}}
\newcommand{\yN}{\overline{\cal N}}\bigskip

\section{Illustrative Numerical Examples}
\label{sec-numres}
In this section, we present illustrative examples of numerical integrations of Eq.(\ref{eq:nueq})
that simulate solutions that are bounded for all real values of the independent variable.\footnote{In 
this section, and in the Appendix, Eq.(\ref{eq:nueq}) will sometimes be written in the form of 
Eq.(\ref{a.1}), and the independent variable $u$ will sometimes be written as $t$.} In 
carrying out these numerical integrations, we have selected functional forms for ${\cal T}_A$,
${\cal T}_C$, and ${\cal N}$ that are consistent with assumptions (\ref{assumA}) to (\ref{assumC}) and that 
include examples of ${\cal N}$  that belong to the trivial homotopy class with $\ell=0$, 
as well as others in which 
$\lim_{t{\rightarrow}\infty}{\cal N}=2\pi\ell$ with $\ell{\neq}0$. The technical considerations on 
which these simulations are based, and the individual examples, constitute the remainder of 
this section
\bigskip

In Theorem \ref{thm1} in the Appendix, we have established the existence of
a solution ${\yN}$ satisfying equation (\ref{a.1}) and
boundary conditions (\ref{a.limit}), provided assumptions
(\ref{assumA}) to (\ref{assumC}) are satisfied.
(It is noted that $w$ and $y$ in the Appendix represent $\wN$ and 
$\yN$, respectively.)
Such assumptions essentially say that $e^t {\cal T}_A$ and
$e^t {\cal T}_C$ behave like 0 for large $|t|$, and that
$\wN$ behaves like 0
as $t \rightarrow - \infty$, and
like $2 \ell \pi$ for some integer
$\ell$, as $t \rightarrow \infty$. \bigskip

Remark 3 in Theorem \ref{thm1} asserts that there are an
infinite number of solutions to Eq.(\ref{a.1}). 
But the question arises: Will numerical solutions of Eq.(\ref{a.1}), that are bounded 
in the entire interval $(-\infty, \infty)$ and normalized so that 
$\lim_{t{\rightarrow}-\infty}\yN=0$, always
settle down to the same two values of $m$ as in the case for the damped pendulum?
(It is well known, in the case of Eq.(\ref{eq:nueqzero}) which describes 
a damped pendulum, that there are exactly two classes of solutions --- 
one for $m=1$ and another for $m=-1$.)
We will carry out some numerical experiments to address this
question, and document some of our results in this section.\bigskip

Since our numerical computations will be done on a finite
interval $[-M,M]$ for some large value $M>0$,
we will first construct the appropriate boundary
condition at $t=-M$. With the solution $\yN$ going
to zero as $t \rightarrow -\infty$, one can impose the boundary condition
$\yN(-M)=0$. However, a better boundary condition can be
constructed, which leads to reduced numerical errors even when we use
only a moderate
value of $M$. This reduced interval size for $[-M,M]$ leads
to more efficient and more accurate calculations.
Such attributes of numerical infinity are well known
(see Chapter 4, Ref. \cite{Keller}).\bigskip

Since both $\yN$ and $\wN$ tend to zero as $t$ approaches $- \infty$, we employ a Taylor's 
expansion to extract the leading
order behavior of the terms in equation (\ref{a.1}) in that regime. We keep only the leading order 
terms in $\wN$ and $\yN$, taking into account that while both are small when $t{\rightarrow}-\infty$, 
the relative magnitudes of $\yN$ and $\wN$ are not known.
In this way, we find that
\begin{equation}
\wN \cos (\yN + \wN) - \sin ( \yN + \wN) \approx
- \yN -  \wN^3/3 \;.
\end{equation}
Hence from (\ref{a.1}), we expect that Eq.(\ref{eq:nueq}) is well approximated by
\begin{equation}
\yN\,''+ \yN\,'- 2 \yN - 2 \wN^3/3 - \beta e^t {\cal T}_A \wN^2/2
- \beta e^t {\cal T}_C \wN  \approx 0
\;.
\label{simp}
\end{equation}
where $'$ designates differentiation with respect to $t$.
The relative magnitudes of the three source terms
containing $\wN$
in the above equation as $t{\rightarrow}-\infty$, will determine which one will dominate over the others.
Assuming that the dominant behavior of these three source terms
is $\gamma e^{\omega t}$ for some constants $\gamma$ and $\omega>0$
as $t$ approaches $- \infty$,  we can further simplify Eq.(\ref{simp}) and
obtain
the following linear non-homogeneous equation with constant coefficients:
\begin{equation}
\yN\,''+ \yN\,'- 2 \yN + \gamma e^{\omega t}
  \approx 0
\;.
\label{approx}
\end{equation}
If we had exact equality in the above equation,
the complementary solutions would be spanned by $e^{-2t}$ and $e^{t}$.
A particular solution of such an equation would be of the form $\gamma_1 e^{\omega t}$,
for some constant $\gamma_1$.
Since $\yN$ is bounded as $t {\rightarrow} -\infty$, we expect solutions for large negative $t$ to 
be of the form
\begin{equation}
\yN \approx C e^{t} + \gamma_1 e^{\omega t}
\end{equation}
for some constant $C$. The following two sets of circumstances may apply:

\vspace{.15in}
\noindent
Case I:  $\omega >1$.

\noindent
In this case, $\yN \approx C e^{t}$. In order to avoid the unbounded complementary solution
$e^{-2t}$, we will employ the boundary condition
$\yN\,'(-M)= \yN(-M) $. We can assign an arbitrary value, $\delta$, to both $\yN$ and to ${\yN}^{\,\prime}$
at $t=-M$, so long as $M$ is sufficiently large and $|\delta|$ is
sufficiently small. Thus we can employ a standard initial value problem-solver
(like Runge Kunta method with local error control, as can be
found in any standard software)
to solve equation (\ref{a.1}), and integrate Eq.(\ref{eq:nueq}) from $t=-M$ to a large positive
 value of $t$. We can also integrate backwards from $t=-M$ to a negative value of $t$ whose absolute 
value is even larger than $M$. 

\vspace{.15in}
\noindent
Case II: $\omega <1$.

\noindent
$\yN \approx C e^{\omega t}$. In order to avoid the unbounded complementary solution
$e^{-2t}$ in this case, we would  have to employ the boundary condition
$\yN\,'(-M)=\omega \yN(-M)$.

\vspace{.15in}
\noindent
In all the numerical experiments presented in the following illustrative examples, 
the dominant behavior as $t{\rightarrow}-\infty$ is that of the complementary
solution $e^t$, so that all our examples belong to case I.
We therefore impose the initial conditions $\yN\,'(-M)=\yN(-M)=\delta$,
for some large $M>0$ and small $|\delta|$ in these experiments.
Provided such conditions are met, when we integrate equation (\ref{a.1})
backward in $t$ towards $- \infty$,
numerical solutions of $\yN$ stay close to
the value 0 for a long interval.
This will be clearly seen in Fig. 1 to Fig. 6 below.
Since we also have an unbounded mode $e^{-2t}$ near $t=-\infty$,
$\yN=0$ acts like a saddle point. Hence if we continue to integrate
backward in time, the numerical solution will eventually blow up,
as one can never get rid of the unbounded mode entirely
in numerical calculations, even though Theorem  \ref{thm1} demonstrates that 
solutions exist that never deviate from 0 as  $t{\rightarrow}-\infty$.\bigskip

Should we use conditions other than $\yN\,'(-M)=\yN(-M)$, the 
magnitude of the solutions
will become large very quickly as we integrate backward in the variable $t$. A larger
value of $M$ is needed to ensure that condition (\ref{a.limit}a) is
satisfied approximately in such a scenario. This larger domain will
increase the computational cost and decrease the accuracy of
the simulation results.\bigskip

From the proof of Theorem \ref{thm1}, we know that for any value
of $\delta$ and $M$, a solution $\yN$ will settle down to a value
as depicted by condition (\ref{a.limit}b). Hence we expect that there are
infinitely many solutions to Eq.(\ref{a.1}) satisfying boundary conditions
(\ref{a.limit}). But the existence of an infinity of solutions does not imply 
that there are infinitely many $m$ values that these solutions approach as 
$t{\rightarrow}\infty$. 
We have carried out numerous numerical experiments to address this question. 
These experiments show that solutions of Eq.(\ref{a.1}) exist, for the identical 
${\cal T}_A$, ${\cal T}_C$, and ${\cal N}$, for which 
$\lim_{t{\rightarrow}-\infty}\overline{\cal N}=0$, which demonstrate that   
$\overline{\cal N}$ approaches limits corresponding to at least three different 
values of $m$ as  $t{\rightarrow}\infty$. We will
document some of the more interesting numerical results below:

\subsection{Experiment 1}
\label{sec-exp1}
In this experiment we have used 
$${\cal N}=\frac{10e^t}{(1+e^{2t})^2}\;\;\;\mbox{and}\;\;\; {\cal T}_A={\cal T}_C=
t^3\exp(-3t^2)\left(1-\exp(-6t^2)\right)$$ 
with $\beta=2gr_0=250$, where we have taken $-M$ as the value of negative numerical infinity with
$M=13.03423$. We have set $\overline{\cal N}=\overline{\cal N}^{\,\prime}=10^{-6}$ at $t=-M$. 
The result, using a standard initial value problem-solver, is represented in Figure 1. We observe that 
$\overline{\cal N}$ is essentially zero from about $t=-5$ to $t=-25$ 
( $|\overline{\cal N}(-25)|{\approx}10^{-5}$),  that it hovers near the 
unstable equilibrium position  $\overline{\cal N}=6\pi$ to within better than 
1 part per $10^5$ at $t=10$, and 
stabilizes to its final position at $\overline{\cal N}=7\pi$. With reference to Eq.(\ref{eq:windnew}),
$\ell=0$ and $m=7$. The winding number for the gauge-invariant gauge field in this 
numerical experiment is $Q=\frac{7}{2}$.

\subsection{Experiment 2}
\label{sec-exp2}
We employ exactly the same data here as in Experiment 1,
except that in our choice of numerical infinity $-M$ in this experiment, $M=13.034$; and that we 
have set $\overline{\cal N}=\overline{\cal N}^{\,\prime}=-10^{-6}$ at $t=-M$.
The result is represented in Figure 2. We observe in this case that 
$\overline{\cal N}$ is again essentially zero from about $t=-5$ to $t=-25$
( $|\overline{\cal N}(-25)|{\approx}10^{-5}$),  that it briefly hovers near the 
unstable equilibrium position  $\overline{\cal N}=6\pi$ in the vicinity of $t=8$, and 
stabilizes to its final position at $\overline{\cal N}=5\pi$. With reference to Eq.(\ref{eq:windnew}),
$\ell=0$ and $m=5$. The winding number for the gauge-invariant gauge field in this 
numerical experiment is $Q=\frac{5}{2}$.

\subsection{Experiment 3}
\label{sec-exp3}
We again employ exactly the same data here as in Experiments 1 and 2,
except that in our choice of numerical infinity $-M$ in this case $M=30$, and that 
we have set 
$\overline{\cal N}=\overline{\cal N}^{\,\prime}=10^{-6}$ at $t=-M$.
The result is represented in Figure 3. We observe in this case that 
$\overline{\cal N}$ is again essentially zero from about $t=-5$ to $t=-25$
( $|\overline{\cal N}(-40)|<2{\times}10^{-10}$),   
that it remains within approximately 1\% of $\pi$ in the vicinity of $t=-9$, and that it
stabilizes to its final position at $\overline{\cal N}=9\pi$. With reference to Eq.(\ref{eq:windnew}),
$\ell=0$ and $m=9$.  The winding number for the gauge-invariant gauge field in this 
numerical experiment is $Q=\frac{9}{2}$.

\subsection{Experiment 4}
\label{sec-exp4}
In this experiment we have used 
$${\cal N}=\frac{100e^t}{(1+e^{2t})^2}+\left(1-\frac{1}{1+e^t}\right)2\pi
\;\;\;\mbox{and}\;\;\; {\cal T}_A={\cal T}_C=
t^3\exp(-3t^2)\left(1-\exp(-6t^2)\right)$$ 
with $\beta=2gr_0=250$, and 
we have taken  $-M$ as the value of negative numerical infinity, with
$M=20$. We have set $\overline{\cal N}=\overline{\cal N}^{\,\prime}=-10^{-4}$ at $t=-M$. 
The result is represented in Figure 4. We observe that 
$\overline{\cal N}$ is essentially zero from about $t=-12$ to $t=-30$ 
( $|\overline{\cal N}(-30)|{\approx}10^{-5}$);  that it briefly hovers near
$\overline{\cal N}=-\pi$ from approximately  $t=-8$ to $t=-6$ to within about .2\%, and that it
stabilizes to its final position at $\overline{\cal N}=29.6874$. Since $\tan^{-1}(2\pi)=1.41297$,
and $29.6874-1.41297=28.2744=9\pi$, this experiment exemplifies an $\ell=1$ and $m=9$ case.
The winding number for the gauge-invariant gauge field in this 
numerical experiment is $Q=5.5677$. \bigskip

\subsection{Experiment 5}
\label{sec-exp5}
We employ exactly the same data here as in Experiment 4, but choose $M=15$.
In this experiment, we
have set $\overline{\cal N}=\overline{\cal N}^{\,\prime}=-10^{-4}$ at $t=-M$.
The result is represented in Figure 5. We observe in this case that 
 $|\overline{\cal N}(-25)|{\approx}10^{-5}$,  and that it
stabilizes to its final position at $\overline{\cal N}=17.1214$. Since $\tan^{-1}(2\pi)=1.41297$,
and $17.1214-1.41297=15.7084=5\pi$, this experiment exemplifies an $\ell=1$ and $m=5$ case. 
The winding number for the gauge-invariant gauge field in this 
numerical experiment is $Q=3.5677$. \bigskip

\subsection{Experiment 6}
\label{sec-exp6}
We employ exactly the same data here as in Experiments 4 and 5, but choose $M=30$.
In this experiment, we
have set $\overline{\cal N}=\overline{\cal N}^{\,\prime}=10^{-4}$ at $t=-M$.
The result is represented in Figure 6. We observe in this case that 
 $|\overline{\cal N}(-35)|{\approx}10^{-6}$, that it  hovers near
$\overline{\cal N}=\pi$ from approximately $t=-17$ to $t=-6$ to about 3\% and that it
stabilizes to its final position at $\overline{\cal N}=35.9699$. Since $\tan^{-1}(2\pi)=1.41297$,
and $35.9699-1.41297
=34.557=11\pi$, this experiment exemplifies an $\ell=1$ and $m=11$ case. 
The winding number for the gauge-invariant gauge field in this 
numerical experiment is $Q=6.5677$.  \bigskip

\subsection{Comments on Numerical Experiments}
\label{sec-comm}
We note that
in all six experiments, we were able to calculate backward in $t$
from $t=-M$ for a long range, and still could obtain small values of $\yN$.
This shows that our choice of initial conditions
$\yN\,'(-M)=\yN(-M)$ has been effective and reliable.\bigskip

With reference to Theorem \ref{thm1}, we note that in Experiments 1, 2 and 3, 
the given $\wN$ leads to
$\ell=0$ for the limit specified in Assumption \ref{assumB}; 
similarly, in Experiments 4, 5 and 6,
the given $\wN$ leads to $\ell=1$. 
Hence,  for large $t$, in Experiments 1,2 and 3, $\yN$ converges to $m \pi$ for some integer $m$,
and, in Experiments 4, 5 and 6, $\yN$ converges to $\tan^{-1}(2\pi)+m \pi$ 
for some integer $m$, 
in accordance with Remark 2 in the Appendix. We observe that for each of two values of $\ell$, 
$m$ is a different integer in each experiment. Thus the 
numerical results show that there are cases in which, 
for the same ${\cal N}$, ${\cal T}_A$, and ${\cal T}_C$, $\overline{\cal N}$ can converge to at 
least three different values of $m$. 
\bigskip

A linearization analysis shows that if $m$ is odd,
then $m \pi$ behaves like a stable equilibrium point. However, when $m$
is even, $m \pi$ acts like a saddle point.
(That is why we have a saddle point behavior at $t=-\infty$, since
it corresponds to $m=0$, which is even.)
A good analogy can again be made to the damped pendulum case.
The even values of $m$ correspond to an inverted pendulum whose equilibrium
is unstable, while
the odd values correspond to the lowest stable equilibrium point.
While it is theoretically possible for a moving pendulum to stop exactly
at the inverted position, any slight error will prevent us from observing
such a phenomenon in numerical calculations. (This is equivalent
to observing that we can never completely eliminate the unbounded mode in numerical
calculations.) However, in principle, there is no reason why $\yN$ 
cannot approach an unstable limit point at which $m$ is even, as $t{\rightarrow}\infty$.
In fact, Experiments 1 and 2 substantiate this claim, since the solutions 
$\overline{\cal N}$ hover near $6\pi$ for such an extended interval.
It is not surprising to find that the precise values of initial conditions that bring 
about the transition from $5\pi$ to $7\pi$ are sensitive to the numerical accuracy of
the computation. In Experiments 1 and 2, a change in the numerical accuracy of the 
computation can require a 1\% change in the value of $M$ to obtain similar qualitative results.

\bigskip

\input{epsf.tex}

\begin{figure}[ht]
\begin{center}
\leavevmode
\epsfxsize=3in
\epsfbox{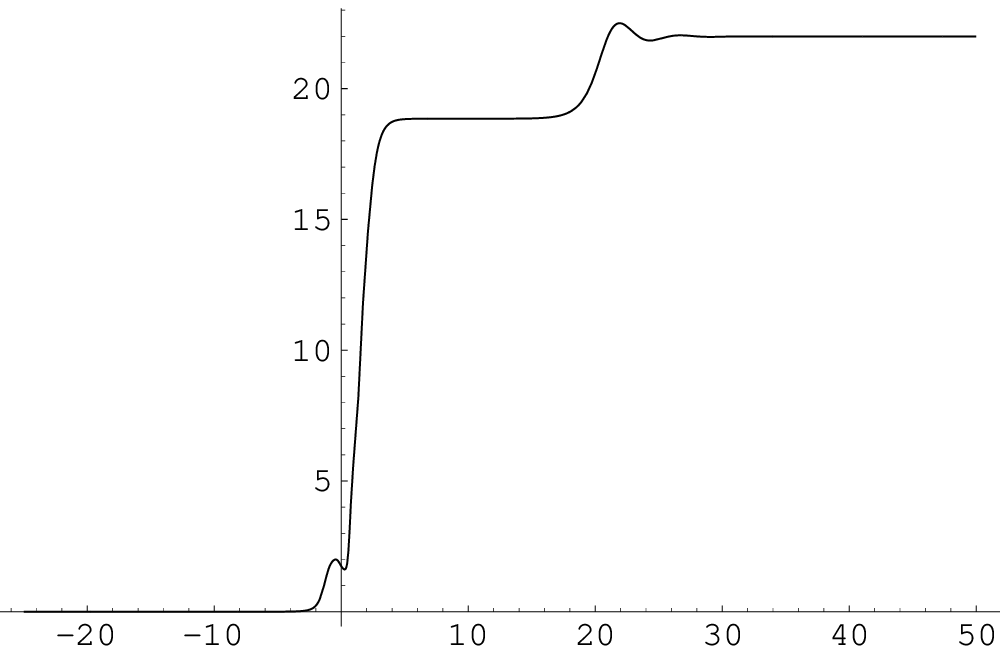}
\end{center}
\caption{\label{fig1}}
\end{figure}

\begin{figure}[ht]
\begin{center}
\leavevmode
\epsfbox{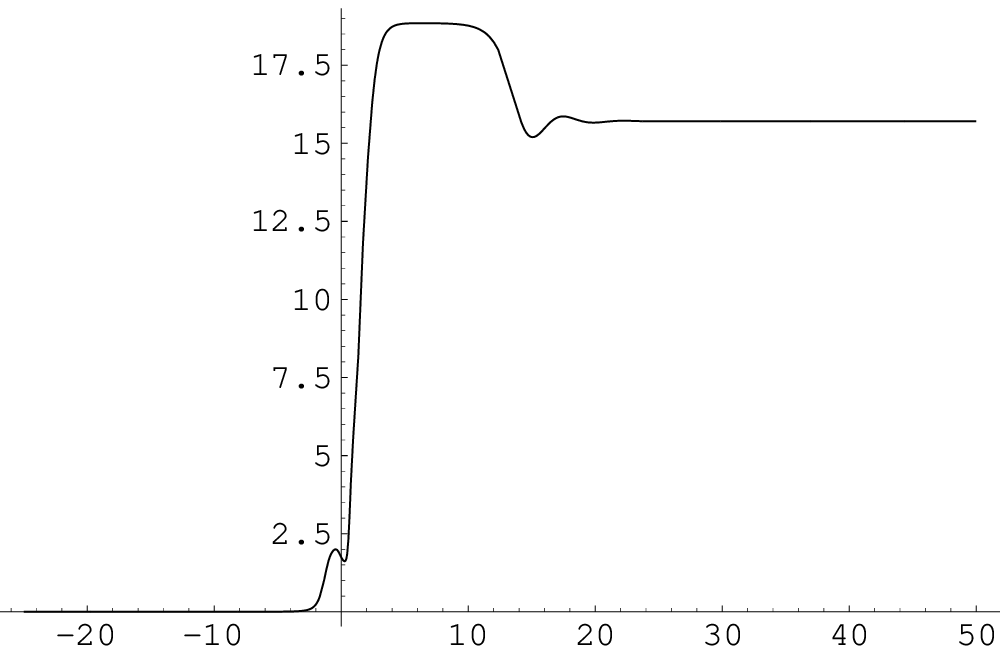}
\end{center}
\caption{\label{fig2}}
\end{figure}

\begin{figure}[ht]
\begin{center}
\leavevmode
\epsfxsize=3in
\epsfbox{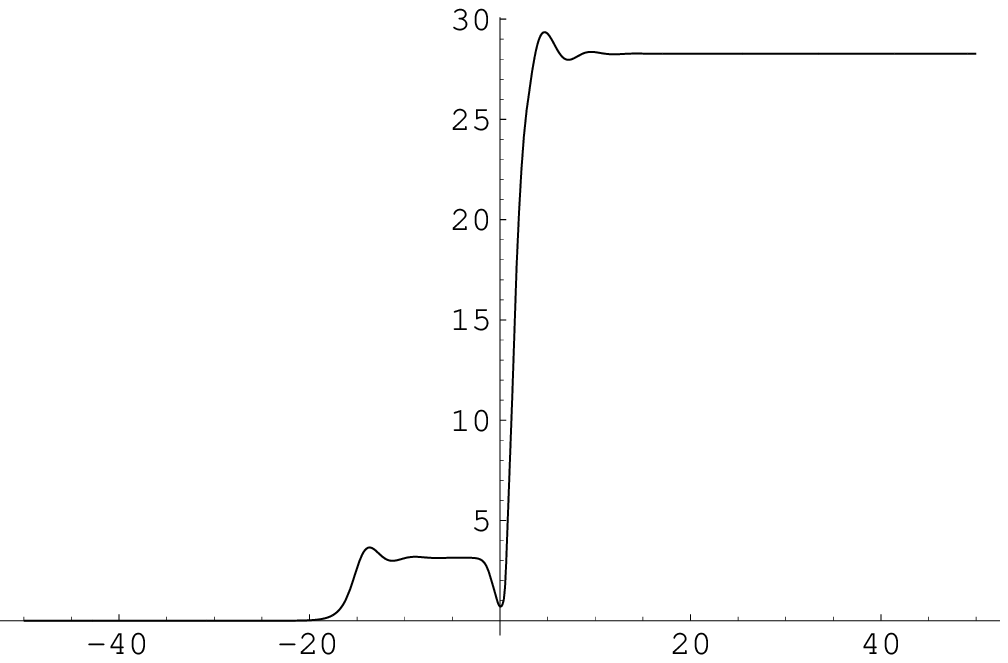}
\end{center}
\caption{\label{fig3}}
\end{figure}

\begin{figure}[ht]
\begin{center}
\leavevmode
\epsfxsize=3in
\epsfbox{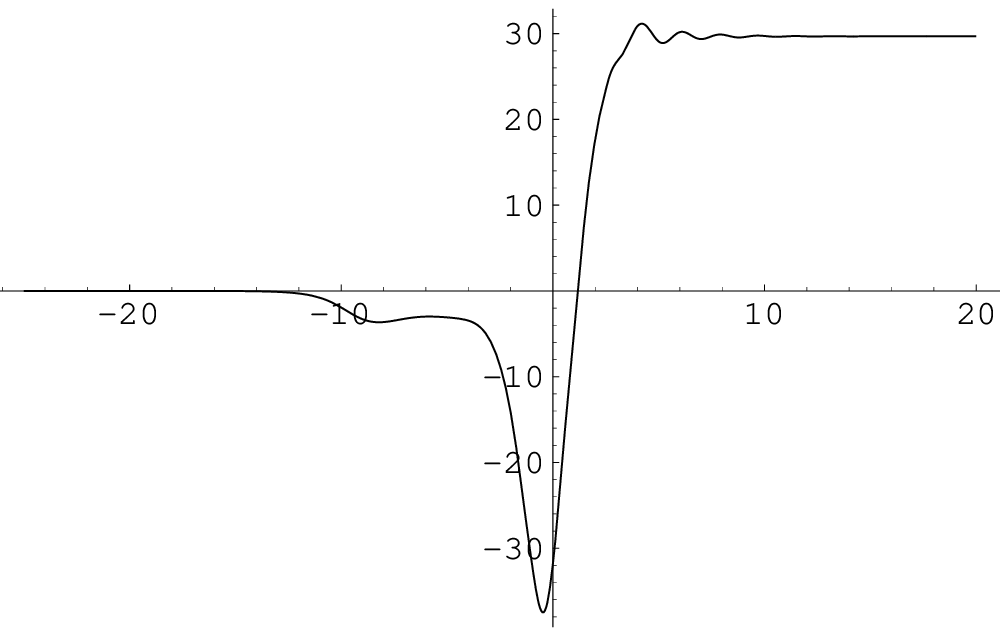}
\end{center}
\caption{\label{fig4}}
\end{figure}

\begin{figure}[ht]
\begin{center}
\leavevmode
\epsfxsize=3in
\epsfbox{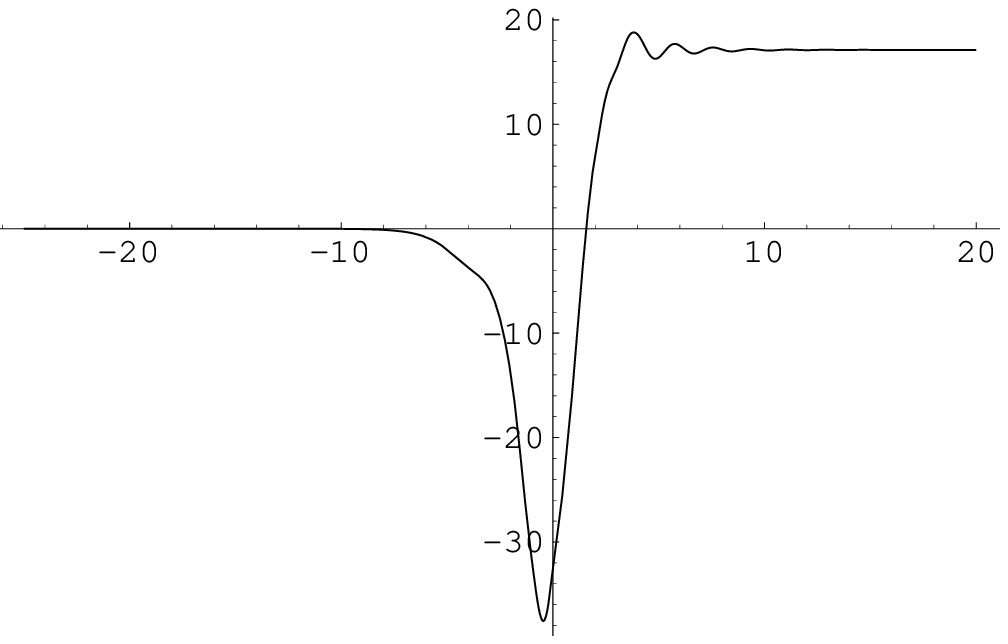}
\end{center}
\caption{\label{fig5}}
\end{figure}

\begin{figure}[ht]
\begin{center}
\leavevmode
\epsfxsize=3in
\epsfbox{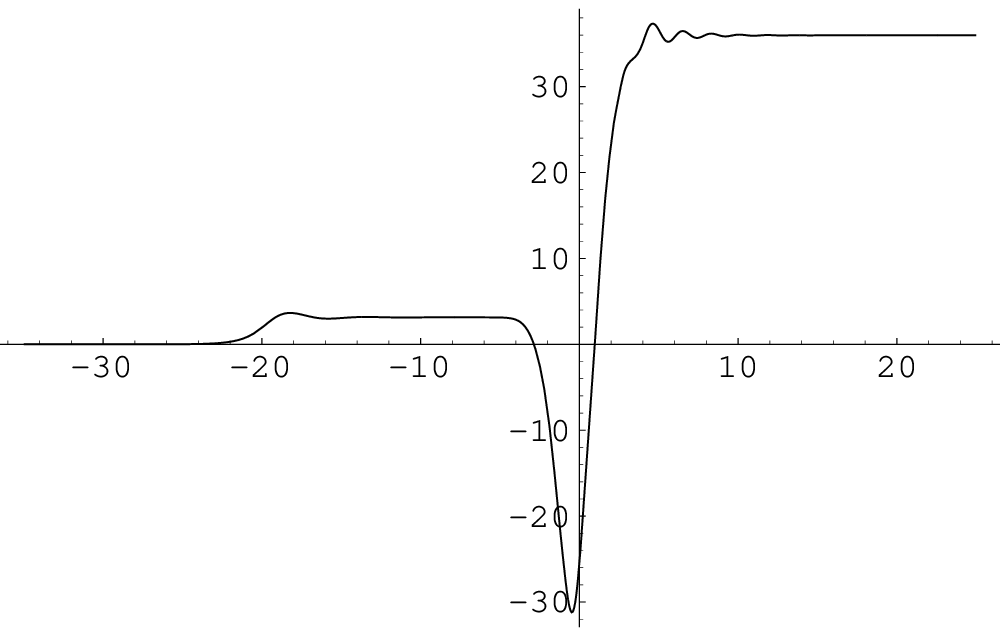}
\end{center}
\caption{\label{fig6}}
\end{figure}

\acknowledgements
One of us (KH) thanks Prof. Gerald Dunne for a number of helpful conversations.
The work of K. Haller and L. Chen was supported by the Department of Energy
under Grant No. DE-FG02-92ER40716.00. The work of Y. S. Choi was 
 partially supported by NIH grant~1P41-RR13186-01.

\appendix
\section{}
In Section \ref{sec-top}, the discussion is based on the properties of the
solutions of Eq.(\ref{eq:nueq}).
The theorem that establishes these solutions is proven in this Appendix. In this
proof, the function
$\overline{\cal N}$ and ${\cal N}$ are represented as $y$ and $w$ respectively,
and the independent
variable $u$ is represented as $t$. \bigskip

Let ${\cal T}_A$,
${\cal T}_C$, and $w$ be given $C^1$ (continuously differentiable) functions
of $t$ in
the interval $(-\infty,\infty)$, let $\beta$ be a given positive constant, and let
$\ell$ be a given integer.
We assume that there is a positive constant $\alpha$ such that \beq
\lim_{t \rightarrow - \infty} \; e^{(1-\alpha)t} {\cal T}_A =0 \;, \;\;\;
\lim_{t \rightarrow -
\infty} \; e^{(1-\alpha)t} {\cal T}_C =0 \;, \;\;\; \lim_{t \rightarrow -
\infty} \;
e^{-\alpha t} w =0 \;, \;\;\; \label{assumA}
\eeq
\beq
\lim_{t \rightarrow \infty} \; e^{(1+\alpha) t} {\cal T}_A =0
\;, \;\;\; \lim_{t \rightarrow \infty} \; e^{(1+\alpha) t} {\cal T}_C =0 \;,
\;\;\;
\lim_{t \rightarrow \infty} \; e^{\alpha t} (w- 2 \ell \pi) =0 \;,
\label{assumB}
\eeq
and
\beq
\lim_{t \rightarrow \infty} \; e^{t} {\cal T}_A' \;, \;\;
\lim_{t \rightarrow \infty} \; e^{t} {\cal T}_C' \;, \;\;\; \mbox{and} \;\;
\lim_{t \rightarrow \infty} \; w'' \; \; \mbox{exist, }
\;\; \lim_{t \rightarrow \infty} \;e^{\alpha t} w'=0 \;. \label{assumC}
\eeq
(i.e. the functions $w$, $e^t {\cal T}_A$ and $e^t {\cal T}_C$ are
exponentially decaying at both positive and negative infinity.) Consider the
equation
\begin{equation}
\begin{array}{lr}
\frac{d^2\,{y}}{dt^2}+\frac{d\,{y}}{dt}+ 2\,\left \{ w \cos (y +w) -
{\sin}({y}+w) \right \} & \\ \\ +\beta e^{t}\left\{{\cal
T}_A({\cos}(y+w)-1)-{\cal T}_C\,
{\sin}(y+w)\right\}& =0.
\end{array}
\label{a.1}
\end{equation}
We have the following theorem.

\begin{theorem}
Under the assumptions (\ref{assumA}) to (\ref{assumC}) on $w$, ${\cal T}_A$ and
${\cal T}_C$ stated above,
there is a solution $y$ to equation (\ref{a.1}) on the interval $(-\infty,
\infty)$. Moreover,
\beq
\mbox{\em (a)}\;\;\lim_{t \rightarrow -\infty} y(t)= 2 n \pi \;, \;\;and\;\;
\mbox{\em (b)}\;\;\;
\lim_{t \rightarrow \infty} y(t) = \tan^{-1}\left(2{\ell} \pi\right)+m\pi \;,
\label{a.limit}
\eeq
where $n$ and m are integers and $\tan^{-1}\,:\,{\bf R}{\rightarrow}\left(-\frac{\pi}{2},
\frac{\pi}{2}\right)$.
\label{thm1}
\end{theorem}

\noindent
{\bf Remark 1:} By using the substitution $Y=y- 2 n \pi$, it suffices to
prove the above theorem when $n=0$. \smallskip

\noindent
{\bf Remark 2:} When $\ell=0$, then equation (\ref{a.limit}b) yields
$\lim_{t \rightarrow \infty} y(t)= m \pi$ for some integer $m$.
When $\ell{\neq}0$, then equation (\ref{a.limit}b) yields
$\lim_{t \rightarrow \infty} y(t)= \tan^{-1}(2{\ell}\pi)+m \pi$ for some integer $m$.
\smallskip

\noindent
{\bf Remark 3:} Actually the proof shows that there are an infinite number
of solutions to our problem.

\vspace{.15in}
\noindent
{\bf Proof:} Because of the Remark 1, we let $n=0$. As $t \rightarrow -\infty$,
we expect that the contributions from the $w$, ${\cal T}_A$ and ${\cal T}_C$
terms are negligible due to the assumptions (\ref{assumA}), so that $y=0$
behaves
exactly like the unstable equilibrium point in the damped pendulum case. We now
prove
that this intuition is indeed what happens. \bigskip

We rewrite the equation as
\[
y''+y'-2y +f(t,y)=0 \;,
\]
where
\[
f(t,y) \equiv 2\left \{(y -\sin (y+w))+w \cos(y+w) \right\}
+ \beta e^{t} \left\{{\cal
T}_A({\cos}(y+w)-1)
-{\cal T}_C\,
{\sin}(y+w)\right\}.
\]
It is noted that $|f(t,0)| \leq o(1)\; e^{ \alpha t}$ and $f_y(t,y)=o(1)$, when
$t$ is
large and negative and $|y|$ is sufficiently small.
Next we convert it to a system of two equations, \[
\begin{array}{rl}
y'& =z \;, \\ \\
z'& = -z +2 y -f(t,y) \;.
\end{array}
\]
If, besides the conditions that $f$ satisfies, $f(t,0)$ also were $0$ for all
$t$,
a standard theorem in ordinary differential equations (theorem 4.1, p.330, Ref.
\cite{Codd} )
would enable us to conclude
that, for any negative $t_0$ with sufficiently large modulus, there is a one
dimensional
manifold $S$, which depends on $t_0$,
in the phase space
$(y,y')$, such that if the initial conditions $y(t_0)$ and $y'(t_0)$ lie on $S$,
then
$y \rightarrow 0$ as $t \rightarrow -\infty$. In our case, we
have $|f(t,0)| \leq o(1)\; e^{\alpha t}$ for large and negative $t$.
The same proof given in theorem 4.1 in Ref. \cite{Codd} still works. The only
alteration needed is in the
first step, when we iterate Eq. (4.11) in that theorem, since $f(t,0)$ is not
zero anymore.
However, because of the exponential decay of $f(t,0)$ for large negative $t$,
Eq. (4.12) in that proof in Ref. \cite{Codd} will still be valid, and the
same proof stands.
Fix $t_0$.
We have thus constructed a solution $y$ to equation (\ref{a.1}) on the interval
$(-\infty, t_0]$ satisfying (\ref{a.limit}a).
(Because of the one dimensional manifold $S$,
this gives rise to Remark 3.) \bigskip

Take any $M>0$.
On any fixed bounded interval $[t_0,M]$, $$
| 2\,\left \{ w \cos (y +w) - {\sin}({y}+w) \right \} +\beta e^{t}\left\{{\cal
T}_A({\cos}(y+w)-1)-{\cal T}_C\,
{\sin}(y+w)\right\}|
$$
is uniformly bounded for all $y$. Thus the solution $y$ can never blow up in
finite time,
and exists for all $t \in (-\infty,\infty)$. \bigskip

To finish the proof of this theorem, it suffices to show that the solution has
property (\ref{a.limit}b). For $\ell=0$,
we expect that this is the case physically, because for large $t$ our equation
behaves
like a damped pendulum moving under the influence of gravity only
--- because of assumptions (\ref{assumB}) and (\ref{assumC}) ---
hence we expect that the solution will settle down in an equilibrium point,
i.e. $y=m \pi$ for some integer $m$.
To prove the property (\ref{a.limit}b) for arbitrary integer values of $\ell$,
take $t_1>0$ sufficiently large so that for $t \geq t_1$, $w \leq e^{-\alpha
t}$,
$e^t |{\cal T}_A| \leq e^{-\alpha t}/2$, and $e^t |{\cal T}_C| \leq e^{-\alpha
t}/2$.
Now for $t \geq t_1$,
using the Cauchy-Schwarz's inequality,
\[
\begin{array}{rl}
& \frac{d}{dt} [\frac{(y')^2}{2} + 2 \cos (y +w) + 2 w \sin (y+w)] \\ \\ =
& y' (y'' +2 w \cos(y+w)- 2 \sin (y+w)) + 2 w w' \cos (y+w) \\ \\ = & -(y')^2 -
\beta \exp(t)\left\{{\cal
T}_A({\cos}(y+w)-1)-{\cal T}_C\,
{\sin}(y+w)\right\} y' \\ \\
& \quad \quad
+2 w w' \cos(y+w)
\\ \\
\leq & -(y')^2 + 2 \beta e^t |y'| (|{\cal T}_A|+|{\cal T}_C|) + 2 |w| \; |w'|
\\ \\
\leq & -\frac{(y')^2}{2} + 4 \beta^2 [ e^t (|{\cal T}_A|+|{\cal T}_C|)]^2 + 2 C
e^{- \alpha t}
\\ \\
\leq & -\frac{(y')^2}{2} + 8 (\beta^2 +C) e^{- \alpha t} \end{array}
\]
for some constant $C >0$.
In other words, if we
define $H \equiv \frac{(y')^2}{2} + 2 \cos (y+w)+ 2 w \sin (y+w) +
\frac{8 (\beta^2 +C) e^{- \alpha t}}{\alpha}$, then \beq
\frac{dH}{dt}
\leq -\frac{(y')^2}{2} \leq 0 \;.
\label{a.energy}
\eeq
Hence $H$ is decreasing in $t$ for $t \geq t_1$, and $H \geq -2 - 2
\|w\|_{\infty}$.
Since $\frac{(y'(t))^2}{2} \leq H(t_1) +2 +2 \|w\|_{\infty}$
for $t \geq t_1$, $|y'|$ is bounded for $t \in [t_1,\infty)$. From the
governing equation (\ref{a.1}), $|y''|$ is then bounded for $t \in
[t_1,\infty)$. \bigskip

Next take the derivate of the equation (\ref{a.1}). From this equation,
 we have $|y'''|$ is bounded for $t \in [t_1,\infty)$ due to assumptions
(\ref{assumC}).
We can now conclude that both $|H'|$ and $|H''|$ are bounded for
$t \in [t_1,\infty)$ by some simple calculations. \bigskip

Since $H$ is monotone decreasing and has a lower bound, there exists a constant
$H_0$ such that
$\lim_{t \rightarrow \infty} H = H_0$.
Together with the bound on $|H''|$, we can conclude (p.116, Ref. \cite{Rudin})
that $\lim_{t \rightarrow \infty} H' = 0$. It can be checked that
\beq
\begin{array}{rl}
\frac{dH}{dt} = &
-(y')^2 - \beta \exp(t)\left\{{\cal
T}_A({\cos}(y+w)-1)-{\cal T}_C\,
{\sin}(y+w)\right\} y' \\ \\
& \;\;
+ 2 w w' \cos (y+w)
- 8 ( \beta^2+C) e^{- \alpha t}
\;.
\end{array}
\label{Hprime}
\eeq
Taking the limit as $t \rightarrow \infty$, we have \beq
\lim_{t \rightarrow \infty} (y')^2 = 0 \; , \label{yprime}
\eeq
which means $y' \rightarrow 0$ as $t \rightarrow \infty$.
Together with $|y'''|$ being bounded, this yields $\lim_{t \rightarrow \infty}
y''= 0 $
(by the same theorem in Ref. \cite{Rudin}).
Finally, taking the limit $t \rightarrow \infty$ in equation (\ref{a.1}), we
obtain the result
\beq
\lim _{t \rightarrow \infty} ( w \cos (y+w) - \sin (y +w) )=0 \;.
\label{a2}
\eeq
This can be reduced to
$\lim _{t \rightarrow \infty}[ 2 \ell \pi \cos y - \sin y] =0$. This is
equivalent to (\ref{a.limit}b).
The proof of the theorem is now complete.

.

\newpage

\begin{center}{FIGURE CAPTIONS}\end{center}
\bigskip

\noindent
Figure 1. A solution of Eq.(\ref{a.1})) is plotted
against the dimensionless
variable $t$ . The input into the equation is given in Subsection \ref{sec-exp1}.
In this case, $\ell$, the winding number of the gauge field ${\cal N}$, is given by  $\ell=0$, 
and the integer that 
characterizes the $\lim_{t{\rightarrow}\infty}\overline{\cal N}$, is $m=7$.
$Q=\frac{7}{2}$ for this case.\bigskip

\noindent
Figure 2. A solution of Eq.(\ref{a.1}) is plotted
against the dimensionless
variable $t$ . The input into the equation is given in Subsection \ref{sec-exp2}.
In this case, $\ell$, the winding number of the gauge field ${\cal N}$, is given by  $\ell=0$, 
and the integer that 
characterizes the $\lim_{t{\rightarrow}\infty}\overline{\cal N}$, is $m=5$.
$Q=\frac{5}{2}$ for this case.\bigskip

\noindent
Figure 3. A solution of Eq.(\ref{a.1}) is plotted
against the dimensionless
variable $t$ . The input into the equation is given in Subsection \ref{sec-exp3}.
In this case, $\ell$, the winding number of the gauge field ${\cal N}$, is given by  $\ell=0$, 
and the integer that 
characterizes the $\lim_{t{\rightarrow}\infty}\overline{\cal N}$, is $m=9$.
$Q=\frac{9}{2}$ for this case.\bigskip

\noindent
Figure 4. A solution of Eq.(\ref{a.1}) is plotted
against the dimensionless
variable $t$ . The input into the equation is given in Subsection \ref{sec-exp4}.
In this case, $\ell$, the winding number of the gauge field ${\cal N}$, is given by  $\ell=1$, 
and the integer that 
characterizes the $\lim_{t{\rightarrow}\infty}\overline{\cal N}$, is $m=9$.
$Q=5.5677$ for this case.\bigskip

\noindent
Figure 5. A solution of Eq.(\ref{a.1}) is plotted
against the dimensionless
variable $t$ . The input into the equation is given in Subsection \ref{sec-exp5}.
In this case, $\ell$, the winding number of the gauge field ${\cal N}$, is given by  $\ell=1$, 
and the integer that 
characterizes the $\lim_{t{\rightarrow}\infty}\overline{\cal N}$, is $m=5$.
$Q=3.5677$ for this case.\bigskip

\noindent
Figure 6. A solution of Eq.(\ref{a.1}) is plotted
against the dimensionless
variable $t$ . The input into the equation is given in Subsection \ref{sec-exp6}.
In this case, $\ell$, the winding number of the gauge field ${\cal N}$, is given by  $\ell=1$, 
and the integer that 
characterizes the $\lim_{t{\rightarrow}\infty}\overline{\cal N}$, is $m=11$.
$Q=6.5677$ for this case.\bigskip

 \end{document}